\documentclass[aps,prd,twocolumn,showpacs,preprintnumbers,amsmath,amssymb]{revtex4}
\usepackage{graphicx}

\begin{document}


\title{\textbf{Gravitational Space Dilation}}


\author{Richard J.\ Cook}
\email[]{richard.cook@usafa.edu}
\affiliation{Department of Physics, U.S. Air Force 
Academy, Colorado Springs, CO 80840-5701}


\date{\today}

\begin{abstract}
We point out that, if one accepts the view that the standard second on an atomic clock is dilated at low gravitational potential (ordinary gravitational time dilation), then the standard meter must also be dilated at low gravitational potential and by the same factor (gravitational space dilation).  These effects may be viewed as distortions of the time and length standards by the gravitational field, and measurements made with these distorted standards can be ``corrected'' by means of a conformal transformation applied to the usual spacetime metric of general relativity.  Because the amount of gravitational time dilation depends on the location of the observer, the ``correction'' of the metric is specific to a particular observer, and we arrive at a ``single-observer'' picture (or SO-picture) of events in a static gravitational field.  The surprising feature of this single-observer picture is a substantial simplification of interpretation and formalism for numerous phenomena in a static gravitational field as compared to the conventional ``many-observer'' interpretation (or MO-picture) of general relativity.
The principle results of the single-observer picture include: 
(1) the speed of light has 
the invariant value $c$ everywhere (as it does 
in the MO-picture), (2) light rays propagate on geodesics of the 
single-observer three-space, (3) the relativistic radar echo delay, when a 
massive body is brought near the radar propagation path, is attributed 
to an increase in the three-space distance between transmitter and 
target, (4) the gravitational bending of light and 
the precession of perihelion are due solely to the curvature of 
the single-observer's three-space, (5) particle motion in a static  
gravitational field is closely analogous to that in Newtonian 
mechanics permitting a clear comparison of classical and relativistic 
motions without approximation, (6) the \emph{exact} field equation for the 
``gravitational potential'' in the single-observer picture is the linear 
Poisson equation of Newtonian gravitation theory, (7) the three-space 
Maxwell equations in the single-observer picture have the same vector forms as in
flat space (not so in the many-observer picture), (8) as a consequence of 
(7), many standard electrodynamic results, such as Ampere's law and 
Faraday's law are valid in the single-observer three-space, (9) a solar-system test of gravitational space dilation is suggested that seems to be within the capability of existing technology, and finally (10) thermal equilibrium in a gravitational field is characterized by uniform temperature in the SO-picture (not so in the standard MO-picture).
\end{abstract}


\maketitle

\section{INTRODUCTION\label{sec:Intro}}

We motivate the viewpoint taken in this paper with the following ``Parable of the self-centered observer."

\begin{quote}
\textit{A ``self-centered'' observer at $O$ in a static gravitational field
\begin{equation}
d\bar{s}^{2}=\bar{g}_{00}c^{2}d\bar{t}^{2}+\bar{g}_{ij}dx^{i}dx^{j},
\label{Metric}
\end{equation}
observes a standard clock at rest at point $P$ ticking slower than his 
own standard clock of identical construction (gravitational time 
dilation).  Being intolerant of other viewpoints, observer $O$ 
concludes that the observer at $P$ uses a faulty time standard and 
that a correction of $P$'s time measurement is required.}

\textit{Observer $O$ also concludes that the length standard used by 
observer $P$ is in error because the standard meter is defined as the 
distance light travels in time $\tau_{m}=(1/299,792,458)\ s$ at the 
defined speed $c$ of exactly $299,792,458\  m/s$, and so an error 
in time measurement by $P$ translates into an error in his length 
standard (the length standard at $P$ is ``too long'', in the view of 
observer $O$, because $P$'s slow 
clock allows light to travel too long a time in defining his standard).}

\textit{A moments thought by observer $O$ convinces him that, if [as implied 
by (\ref{Metric})], 
\begin{equation}
\mathcal{R}= \left[ \frac{\bar{g}_{00}(P)}{\bar{g}_{00}(O)} \right] ^{1/2}
\label{Rate}
\end{equation}
is the rate of $P$'s clock as measured with $O$'s time standard, then 
the ``incorrect'' time and length intervals ($d\bar{\tau}$ and 
$d\bar{\ell}$) 
measured by $P$ using ``faulty'' local standards are related to the 
``true'' length and time intervals ($d\tau$ and $d\ell$) 
at $P$ by equations
\begin{subequations}
\label{ScalingLaws}
\begin{eqnarray}
d\tau&=&\frac{d\bar{\tau}}{\mathcal{R}}
\label{timescale} \\
d\ell&=&\frac{d\bar{\ell}}{\mathcal{R}}.
\label{lengthscale}
\end{eqnarray}
\end{subequations}
Equation (\ref{timescale}) expresses the well-tested gravitational 
time-dilation effect.
Observer $O$ concludes that both time \emph{and} length are dilated 
at $P$, and by the same factor.  He finds this result satisfying in a 
relativistic theory where space and time are ``made of the same 
stuff'' and therefore ought to be affected similarly by a gravitational field. 
From this point on, he speaks not of time dilation alone but 
of space dilation as well, or of ``spacetime dilation''.}
\end{quote}

The scale change described by equations (\ref{ScalingLaws}) 
 is represented in spacetime notation by the 
conformal transformation $ds^{2}=d\bar{s}^{2}/\mathcal{R}^{2}$ of 
metric (\ref{Metric}).  Under this transformation the proper time 
$d\bar{\tau}$ [$= (-d\bar{s}^{2}/c^{2})^{1/2}$] of observer $P$ goes over into 
the proper time $d\tau$ [$=(-ds^{2}/c^{2})^{1/2}$] of 
observer $O$ as in Eq.~(\ref{timescale}), and the proper distance $d\bar{\ell}$ 
($d\bar{\ell}^{2}=\bar{g}_{ij}dx^{i}dx^{j}$) of observer $P$
goes over into the proper distance $d\ell$ of observer $O$ in accordance with 
Eq.~(\ref{lengthscale}).  Therefore the ``correct'' metric in the view of our self-centered observer (``corrected'' for gravitational time dilation) reads
\begin{eqnarray}
ds^{2}&=& d\bar{s}^{2}/\mathcal{R}^{2} \nonumber \\
 &=& \frac{\bar{g}_{00}(P)c^{2}d\bar{t}^{2}}{[\bar{g}_{00}(P)/\bar{g}_{00}(O)]}
 +\frac{\bar{g}_{ij}dx^{i}dx^{j}}{\mathcal{R}^{2}} \nonumber \\
  &=& -c^{2}dt^{2}+g_{ij}dx^{i}dx^{j},
  \label{SOtrans}
  \end{eqnarray}
where we have trivially rescaled the coordinate time by a constant 
  factor depending on $O$'s location 
  ($t=[-\bar{g}_{00}(O)]^{1/2}\bar{t}$), and have identified
  \begin{equation} 
  g_{ij}=\bar{g}_{ij}/\mathcal{R}^{2}
  \label{spacemetric}
  \end{equation}
  as the spatial metric tensor observer $O$ would regard as correct ($d\ell^{2}=g_{ij}dx^{i}dx^{j}$ is the appropriate spatial metric in observer $O$'s opinion).
    
  The usual spacetime picture in general relativity (the one based on 
  metric $d\bar{s}^{2}$) is operationally founded on the notion of a large 
  number of observers distributed over space; each making local 
  measurements with locally defined standards of time and length 
  (the ``ten thousand local witnesses'' in the words of Taylor and 
  Wheeler) \cite{Taylor}.  We 
  shall refer to this as the ``many-observer picture'' (MO-picture) and the 
  associated metric $d\bar{s}^{2}$ as the metric of \emph{many-observer 
  spacetime} (or 
  \emph{MO-spacetime} for short), and we shall call the spatial 
  section with metric $d\bar{\ell}^{2}=\bar{g}_{ij}dx^{i}dx^{j}$ 
  \emph{many-observer  
  space} (or \emph{MO-space}).  Similarly we shall refer to the picture 
  based on the metric $ds^{2}$ as the \emph{single-observer 
  picture} (or \emph{single-observer spacetime}, or 
  \emph{SO-spacetime}), and we shall call the spatial section with 
  metric $d\ell^{2}=g_{ij}dx^{i}dx^{j}$  
  \emph{Gaussian space} (or \emph{SO-space or G-space}).  The reason for calling 
  the single-observer space ``Gaussian space'' will become apparent 
  in Sec.~\ref{sec:LightRays}  [We are using the over-bar to denote the standard 
  MO-picture variables and no over-bar for SO-picture variables 
  because most of the equations to follow are in the SO-picture and 
  the notation is thereby simplified.  We use the same (unbarred) 
  spatial coordinates $x^{i}$ in both pictures.].
  
  One purpose of the present paper is to show that gravitational space 
  dilation, like gravitational time dilation, is measurable (at least 
  in principle) and that the relativistic radar echo delay experiment may be interpreted as a test of gravitational space dilation.  A second purpose of this paper is to show that the 
  single-observer picture is convenient for the calculation and 
  interpretation of various results in general relativity.  But 
  before doing so, we address an objection that may already be in the 
  mind of the reader.

  Generally speaking, measurement from a distance is problematic in general 
  relativity.  Therefore we should like to emphasize that the single-observer 
  picture does not involve actual measurement from a 
  distance.  In a static gravitational field, regular light pulses 
  from a stationary beacon cross any stationary point at the 
  same rate they are emitted (in coordinate time) because each pulse 
  traverses an identical time-independent field.  Therefore slave clocks distributed 
  at rest over space and ticking each time they receive a beacon pulse 
  from $O$ all tick at the same rate (no gravitational slowing), and 
  any one of them can be used to reproduce $O$'s time and 
  length standards at its location.  Observer $O$'s time unit 
  transferred to $P$ is the time between ticks on the slave clock at 
  $P$, and observer $O$'s length standard transferred to $P$ is the 
  distance light travels at $P$ in time $\tau_{m}$ (=1/299,792,458  
  seconds) \emph{as read on the 
  slave clock at $P$}.  The metric $ds^{2}$ describes the 
  results of measurements made locally with the slave-clock 
  standards, whereas the metric $d\bar{s}^{2}$ describes the results of 
  measurements made with time and length standards based on local
  freely-running, gravitationally-slowed clocks.  Strictly speaking, 
  this is what we mean when we say that $O$ applies his own time and 
  length standards at $P$.  It is the same logic used in comparing 
  clock rates in the gravitational time-dilation experiment.  
  Therefore, no 
  actual measurement from a distance is involved, though, for 
  convenience, we shall often speak loosely in such terms.
  
  Incidentally, the slave clock readings determine a time at each 
  point of space.  If the slave clocks are synchronized by the 
  Einstein synchronization procedure, then they read the coordinate 
  time $t$ ($=[-\bar{g}_{00}(O)]^{1/2}\bar{t}$) introduced above for the 
  metric $ds^{2}$.  This is an operational definition of 
  the time $t$.
  
 The metrics $d\bar{s}^{2}$ and $ds^{2}$ 
  are physically equivalent and conformally equivalent. 
   They represent the same spacetime 
  geometry measured with different standards of length and time 
  \cite{Remark}.  Nevertheless, we shall refer to the different 
  picture as ``different spacetimes'' because ``invariants'' such as 
 the curvature scalars $\bar{R}$ and 
  $R$ have different values in the two pictures and the metrics 
   $d\bar{s}^{2}$ and $ds^{2}$ are not related by a coordinate 
   transformation.
  
  One important result which is the same in both the single-observer 
  and many-observer pictures concerns the 
  local speed of light.  In MO-spacetime the local speed of light is 
  everywhere $c$: $d\bar{\ell}/d\bar{\tau}_{0} =c$ or $d\bar{s}=0$ (here 
  $d\bar{\tau}_{0}$ is the proper time 
  on a local freely-running rest clock).  Under the transformation 
  (\ref{timescale})-(\ref{lengthscale}) [or (\ref{SOtrans})] the local 
  light speed retains this value: 
  $d\ell/d\tau_{0}=(d\bar{\ell}/\mathcal{R})/(d\bar{\tau}_{0}/\mathcal{R}) 
  =d\bar{\ell}/d\bar{\tau}_{0} =c$ (or $ds=0$), where now 
  $d\tau_{o}=dt$ is local slave-clock time, or 
  ``observer $O$'s time'' when we speak loosely.  Hence we have
  
  \begin{quote}

  \textbf{Result I:} \textit{In the single-observer picture the speed 
  of light $d\ell/dt$ has the invariant value $c$ everywhere and at 
  all times.}

  \end{quote}

  The same argument shows that \emph{any} speed has the same value in MO-space 
  and in SO-space (or Gaussian space), i.e., $v=\bar{v}$..  Note 
  that, according to metric (\ref{SOtrans}), the 
  single-observer rest-clock proper time $\tau_{0}$ is the same 
  as the coordinate time $t$, and we use the two interchangeably.

  In Sec.~\ref{sec:LightRays} we show that light rays follow geodesics of the 
  single-observer three space (geodesics of Gaussian space) and discuss some 
  simple consequences of this result.  Sec.~\ref{sec:RadarDelay} addresses the radar 
  propagation problem, and concludes that radar from $O$ directly 
  measures the single-observer distance $\int d\ell$, 
  and the relativistic radar echo 
  delay experiment \cite{Shapiro} may be interpreted as an increase in 
  this 
  distance as the solar mass comes near the radar propagation path.  
  Sec.~\ref{sec:ParticleMotion} treats particle motion in a static gravitational field. 
   Here we find a close analogy between 
  particle motion in Gaussian space and that in Newtonian 
  gravitational mechanics. This 
  permits a clear comparison of classical and relativistic motions 
  \emph{without introducing approximations}, e.g., perihelion 
  precession is attributed solely to the curvature of Gaussian space in the 
 SO-picture.  In Sec.~\ref{sec:GravitationalPotential}, we 
  find that the gravitational acceleration of a particle  
  at rest in Gaussian space, $\mathbf{g}_{0}=-\nabla \Phi$, is determined 
 by the linear Poisson equation 
  $\nabla^{2}\Phi=-4\pi G\rho_{g}$ of Newtonian gravitation theory
  (here seen to be an exact relativistic result)---the only 
  difference between the Newtonian and relativistic gravitation 
  theories being the non-Euclidean geometry of the Gaussian three-space and an 
  energy dependent factor $(\bar{m}c^{2}/E)^{2}$ in the gravitational 
  acceleration $\mathbf{g}=-(\bar{m}c^{2}/E)^{2}\nabla \Phi$ when the 
  particle velocity is nonzero.  In Sec.~\ref{sec:Electrodynamics}, the three-space Maxwell 
  equations are derived for Gaussian space and are found to have the same 
  vector forms as in flat space (not so in the MO-picture).  This leads to a 
  number of electromagnetic theorems, e.g., Amperes law 
  and Faraday's law, that hold in Gaussian  space but do not hold in the usual 
  MO-picture. In Sec.~\ref{sec:DilationTest} we suggest a solar-system test of 
  gravitational space dilation involving radar measurements.  Sec.~\ref{sec:Applications} 
  presents a number of applications of the space-dilation formalism 
  chosen to illustrate how very different the physical 
  interpretations of phenomena can be in the MO- and SO-pictures.  
  The paper concludes in Sec.~\ref{sec:Conclusion} with a summary of results.

\section{\label{sec:LightRays}LIGHT RAYS AS SPATIAL GEODESICS}

Fermat's principle in MO-space \cite{Misner} states that, in a static gravitational field, the 
path taken by light between points $P_{1}$ and $P_{2}$ is the one that 
minimizes the \emph{coordinate time} elapsed in propagating between the two 
points,
\begin{equation}
	\delta\int_{P_{1}}^{P_{2}}d\bar{t} = 0 .
	\label{timeInt}
\end{equation}
Using the null condition $d\bar{s}^{2}= 0$ and (\ref{Rate}) in Eq.~(\ref{Metric}), we find 
that (\ref{timeInt}) can be written as
\begin{equation}
\delta\int_{P_{1}}^{P_{2}}\left[\frac{\bar{g}_{ij}}{\mathcal{R}^{2}} \frac{dx^{i}}{dq}
\frac{dx^{j}}{dq} \right]^{1/2} dq = 0 
	\label{intmod}
\end{equation}
for any parameterization $x^{i}(q)$ of the path between $P_{1}$ and 
$P_{2}$. But Eq.~(\ref{intmod}) is precisely the condition for the 
single-observer path length between the two points to be minimum,
\begin{equation}
	\delta\int_{P_{1}}^{P_{2}} d\ell = 0,
	\label{spaceint}
\end{equation}
where $d\ell^{2} = g_{ij}dx^{i}dx^{j}$  is the spatial part of 
the single-observer metric (\ref{SOtrans}).

\begin{quote}

\textbf{Result II:}\textit{\ Light rays in a static gravitational field are geodesics 
of the single-observer  spatial metric 
$d\ell^{2} = g_{ij}dx^{i}dx^{j}$, i.e., geodesics of Gaussian space.}

\end{quote}

This result can be understood by noting that, in a space where the light speed 
$d\ell/dt = c$ is everywhere the same \emph{and the 
differential $dt$ is exact} (so that the speed $d\ell/dt$ is 
integrable), a principle of least time is equivalent to 
a principle of least distance.  This does not hold true in MO-space where, although 
each local observer measures the same light speed 
$d\bar{\ell}/d\bar{\tau}_{0} = c$, the clocks of different observers run at 
different rates (the local proper time differential $d\bar{\tau}_{0}$ 
is not exact) and it cannot be concluded that the MO-space 
distance is least.  Indeed, it is well known that light rays \emph{do 
not} travel the geodesics of the spatial metric 
$d\bar{\ell}^{2}=\bar{g}_{ij}dx^{i}dx^{j}$ of MO-space.

Let us apply Result II to the Schwarzschild line element
\begin{equation}
	d\bar{s}^{2}=-(1-2m/r)c^{2}d\bar{t}^{2}+\frac{dr^{2}}{(1-2m/r)}
	+r^{2}d\Omega^{2}.
	\label{eq.Scharz}
\end{equation}
where $d\Omega^{2}=d\theta^{2}+sin^{2}\theta d\phi^{2}$.
The rate of a clock at rest at $P$ (coordinates $r, \theta, \phi$) as 
viewed by an observer $O$ at $r_{0}= \infty$ [Eq.~(\ref{Rate})] is
\begin{equation}
	\mathcal{R}=\left(1-\frac{2m}{r} \right)^{1/2} .
	\label{SchRate}
\end{equation}
Hence the spatial part of the single-observer metric 
(\ref{SOtrans}), specialized to the $ \theta = \pi /2 $ plane, reads
\begin{equation}
	d\ell^{2}=\frac{1}{(1-2m/r)} \left[
	\frac{dr^{2}}{(1-2m/r)} +r^{2}d\phi^{2} \right] .
	\label{3Dmetric1}
\end{equation}
Using $\phi$ as the parameter for the trajectory $r(\phi)=1/u(\phi)$, 
we easily obtain the two-space geodesic equation in this Gaussian 
space:
\begin{equation}
	\frac{d^{2}u}{d\phi^{2}} +u=3mu^{2}.
	\label{orbit.eq}
\end{equation}
This is the familiar trajectory equation for light in the 
gravitational field of the sun (in which case $m=1.48\ km$), here 
obtained as a spatial geodesic of G-space.  The 
complete description of light propagation for the single observer 
consists of the trajectory $r(\phi)$ together with  
Result I stating that light 
travels this trajectory with speed $d\ell/dt = c$.  (Here the 
single-observer time $t$ is just the coordinate time $\bar{t}$ 
because the observer is at infinity.).

  The \emph{interpretation} of light deflection is simple in 
Gaussian space.  According to Result II, light is 
bent by a static gravitational field solely because it is traveling the 
shortest path in the non-Euclidean G-space with metric
\begin{equation}
	d\ell^{2}= \frac{1}{(1-2m/r)}
	\left[\frac{dr^{2}}{(1-2m/r)}+
	r^{2}d\Omega^{2}\right] ,
	\label{3metric}
\end{equation}
where $d\Omega^{2}=d\theta^{2}+ sin^{2}\theta d\phi^{2}$  (Here we are not restricted to the $\theta = \pi /2$ plane.)

\begin{quote}

\textbf{MO-Picture Interpretation:}  The interpretation of light deflection in MO-spacetime is more 
complicated. In several books on general relativity we find diagrams 
suggesting that gravitational light deflection may be attributed to the 
non-Euclidean nature of three-space, but we are often not 
warned that ``the light ray is not a geodesic line in 
three-dimensional space'' (the quote is from Pauli \cite{Pauli}).  In 
Will \cite{Will1} we correctly learn that half of the light deflection 
may be attributed 
to the non-Euclidean MO-space and half to 
the principle of equivalence applied in a sequence of freely falling 
reference frames along the ray, as first employed by Einstein 
\cite{Einstein1}.

\end{quote}

We now see that \emph{light rays as spatial geodesics} is a 
valid concept provided we adopt the ``self-centered'' view of the 
single observer who uses his own length standard for distance 
measurement throughout space in place of the locally defined standards.

A rewording of Result II states that the 
three-velocity $c^{k}=dx^{k}/dt$ of a photon (the tangent vector 
to the light ray) undergoes parallel transport in G-space.  This 
follows from Result II by taking $q=t$ as the affine parameter for the 
ray $x^{k}(t)$.  The three-space geodesic equation 
[$d^{2}x^{k}/dt^{2}+\Gamma^{k}_{ij}(dx^{i}/dt)(dx^{j}
/dt) = 0$] is then the equation of parallel transport
 [$dc^{k}/dt = -\Gamma^{k}_{ij}c^{i}dx^{j}/dt$] for 
 the photon three-velocity $c^{k}$ in G-space, and the parallel transport of 
 $c^{k}$ leaves the magnitude of this three-vector invariant at value $c$ 
 (Result I).

\begin{figure}
\includegraphics[width=1.75in]{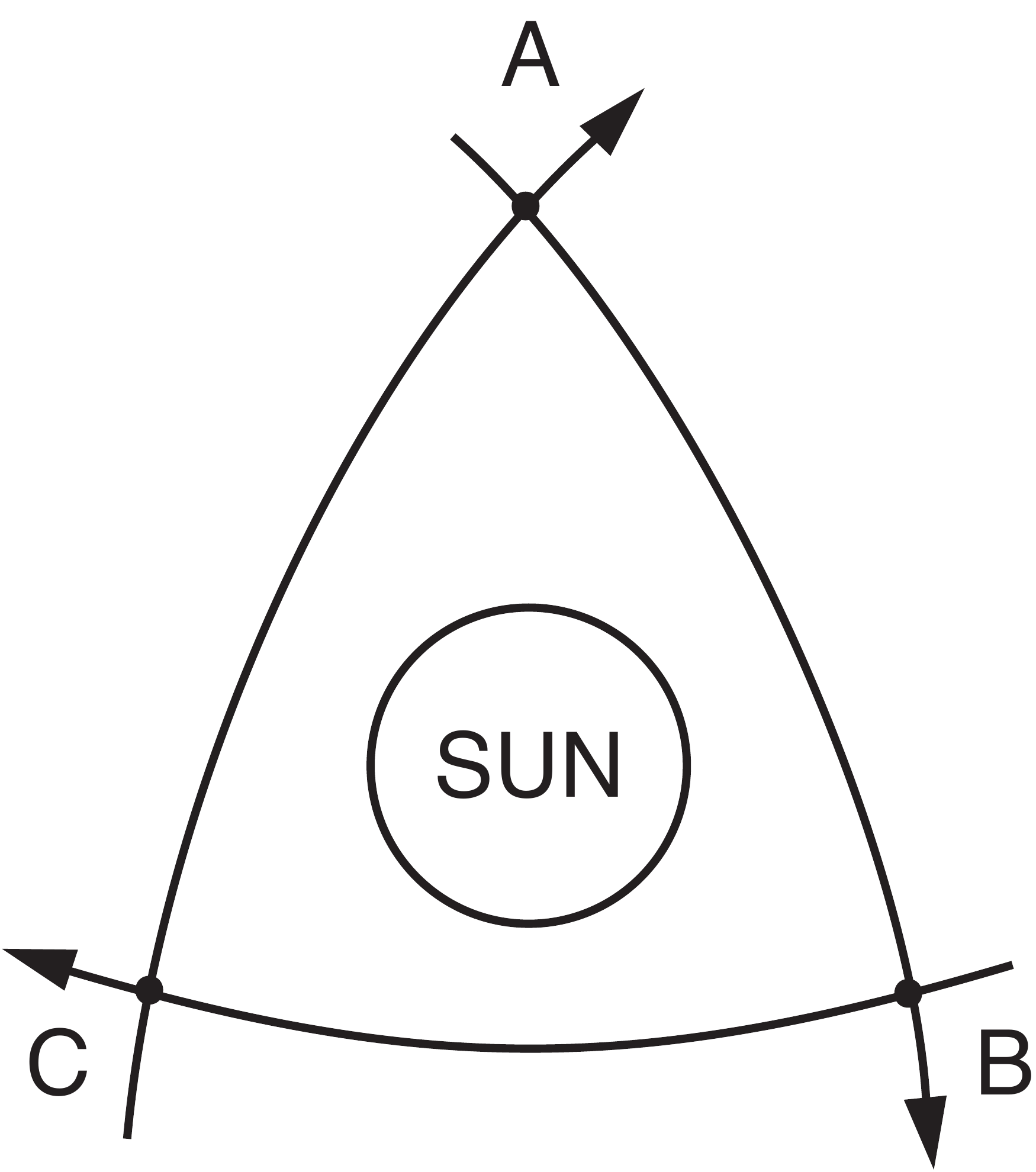}
\caption{\label{fig1}Three gravitationally bent light rays form a 
triangle $ABC$ about the sun.  The sum of the interior angles is 
greater than two right angles.  This shows directly that the 
single-observer three-space (G-space) is non-Euclidean, because the 
light rays are geodesics of G-space.}
\end{figure}

Because the geodesics of G-space are the paths of light rays, the 
non-Euclidean features of this space are measurable optically.  
Fig.~\ref{fig1} shows three intersecting light rays, each bent by the sun's 
gravitational field. 
The interior angles of the triangle formed by these rays sum to 
greater than $\pi$ radians; an earmark of non-Euclidean geometry.  
Therefore,  the measured deflection of light by the sun's 
gravitational field is \emph{direct} evidence for the curvature of 
G-space (It is not \emph{direct} evidence for a non-Euclidean MO-space, 
because light rays are not geodesics of MO-space).  

In Euclidean geometry, a ``biangle'' is a rather uninteresting closed 
figure with two equal sides, two equal angles ($\alpha = \beta = 0$), 
and zero area (two overlapping line segments).  However, in non-Euclidean geometry,
the biangle can open into a 
non-trivial figure with finite area and finite vertex angles.  The 
``football shaped'' figure formed by two lines of longitude on a 
sphere meeting at the poles is an example of a non-trivial 
biangle; an ``equilateral biangle''.

\begin{figure}
\includegraphics[width=3in]{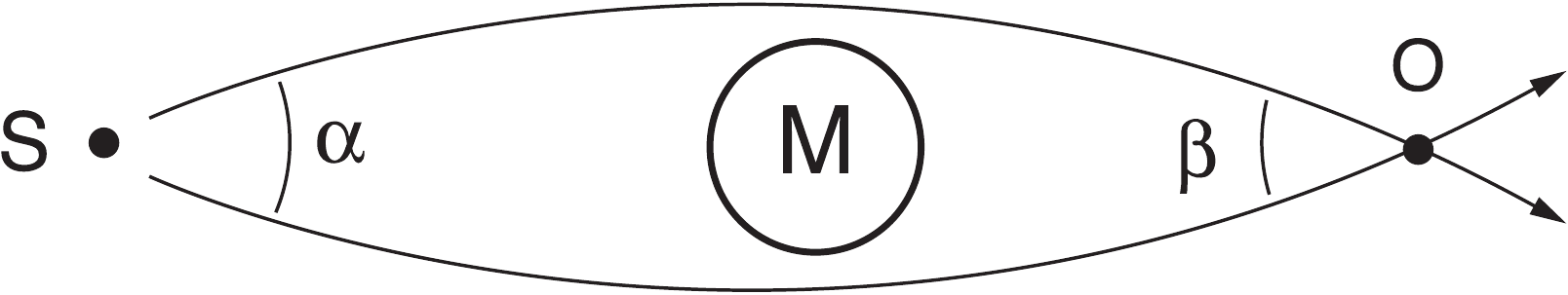}
\caption{\label{fig2}Two light rays diverging from source $S$ are bent 
by the gravitational field of mass $M$ and intersect at observer $O$.  
The closed two-sided figure with vertex angles $\alpha$ and $\beta$ is 
a non-trivial ``biangle'' in the G-space of observer $O$.   The existence of a 
non-trivial biangle (geodesics initially diverging from a point and 
then coming together) implies that this G-space is non-Euclidean.  It 
follows that the double images of a quasar produced by gravitational 
lensing offer \emph{direct} evidence for a non-Euclidean G-space.}
\end{figure}

Two light rays diverging from a point and brought back together by the
gravitational attraction of a mass $M$, as in Fig.~\ref{fig2}, form a biangle in 
the G-space of observer $O$.
  
The existence of a biangle with non-zero 
vertex angles implies that the G-space around mass $M$ is 
non-Euclidean.  We may conclude that the double (or 
multiple) images of quasars, formed by gravitational lensing 
also provide \emph{direct} evidence for a non-Euclidean G-space geometry.

Probably the first attempt to detect non-Euclidean features of 
three-dimensional space was that of 
Carl Friedrich Gauss.  It is 
reported that Gauss used light beams between mountain tops in an 
attempt to detect a deviation from the Euclidean theorem stating that the sum 
of the interior angles of a triangle equals two right angles 
\cite{Bonola}. In this experiment Gauss found it natural to assume that light 
propagates along geodesics of the three-space.  Because the geodesic 
character of light rays is the central feature of the single-observer 
three space, it seems natural to refer to this space as ``Gaussian 
space'' (or ``G-space'' for short) as we have been doing already for a 
while now.

We note in passing that, when a carpenter looks along the edge of a piece of lumber to see if it is ``straight,''  he is using the Gaussian-space definition of ``straight,'' namely the path of a light ray as the straightest possible curve in three-dimensional space.

\section{\label{sec:RadarDelay}RADAR ECHO DELAY}

Consider the G-space distance $\ell$ measured along a 
light ray from the observer at $O$ to the point $P$.  Because the 
single observer at $O$ sees the same light speed $d\ell/dt = c$ 
everywhere (and because the differential $dt$ is exact, i.e., the 
derivative $d\ell/dt$ is integrable), it follows that 
the distance $\ell$ is traversed in time $\Delta t = 
\ell / c$.  If light travels from $O$ to $P$ and back to 
$O$, the total elapsed time on the clock at $O$ is given by 
$\Delta t = 2\ell /c$, or $\ell = c \Delta t/2$, namely the 
radar range formula.  We thus arrive at 

\begin{quote}

\textbf{Result III:}\textit{\ The G-space distance measured along a light ray 
from observer $O$ to point $P$ (the geodesic distance determined by 
the spatial metric $d\ell^{2}= g_{ij}dx^{i}dx^{j}$) is the 
radar distance from $O$ to $P$ measured using a radar unit at $O$.}

\end{quote}

This is a useful observation.  It tells us that, in a 
static gravitational field, when a single observer chooses length and 
time standards throughout space consistent (in his view) with his own 
local standards, his natural measure of distance is 
radar distance.  This singles out radar distance as more natural for 
the single observer than the other possible distance measures 
(MO-space proper distance, 
luminosity distance, angular diameter distance, etc.). By the same 
token, radar is the natural tool for measuring the G-space geometry, 
since it measures G-space distance directly, at least along light rays 
from $O$ (actually the G-space metric for observer $O$ is measured
by radar from any point and in any direction so long as the radar at 
that point uses the local slave-clock time $t$ to measure the radar echo delay).

The relativistic radar echo delay has a simple interpretation 
in G-space.  Consider the Schwarzschild geometry in isotropic 
coordinates:
 \begin{eqnarray}
 	d\bar{s}^{2}&=&-\left(\frac{1-m/2\rho}{1+m/2\rho}\right)^{2}c^{2}d\bar{t}^{2} \nonumber \\
	& &+\left(1+m/2\rho\right)^{4}\left(dx^{2}+dy^{2}+dz^{2}\right).
 	\label{isoScharz}
 \end{eqnarray}
 Here $x, y, z$ are rectangular coordinate markers and
 $\rho = \left(x^{2}+y^{2}+z^{2}\right)^{1/2}$.
 
 To an observer at infinity ($\rho_{0}= \infty$), a stationary clock 
 at radial coordinate $\rho$ runs at the rate
 \begin{equation}
  	\mathcal{R} = \frac{\left(1-m/2\rho\right)}{\left(1+m/2\rho\right)}.
  	\label{isorate}
  \end{equation}
  Hence the G-space geometry is described by the metric
  \begin{equation}
   	d\ell^{2}= \frac{d\bar{\ell}^{2}}{\mathcal{R}^{2}}=\frac{(1+m/2\rho)^{6}}{(1-m/2\rho)^{2}}
   	\left(dx^{2}+dy^{2}+dz^{2}\right).
   	\label{iso3D}
   \end{equation}
   
\begin{figure}
\includegraphics[width=3in]{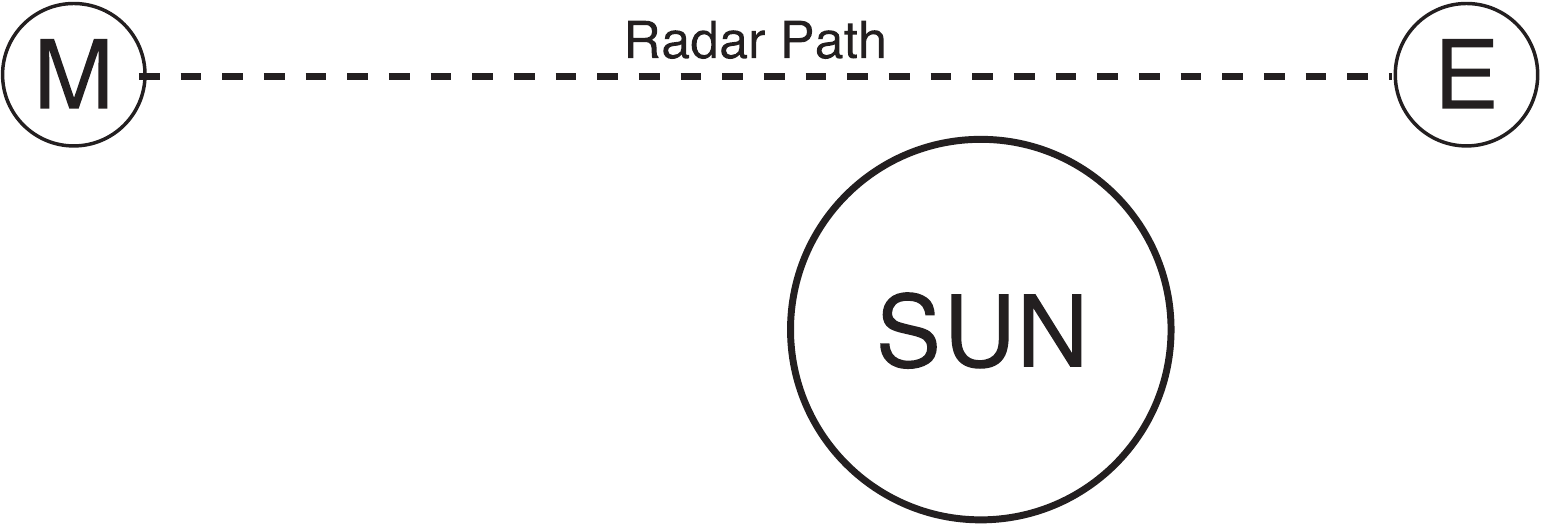}
\caption{\label{fig3}Radar propagates from Earth (E) to Mars (M) and 
back.  The radar path length in Gaussian space is longer when the sun 
is near the propagation path (superior conjunction) than when the sun 
is far removed from the propagation path.  This accounts for the 
relativistic radar echo delay.}
\end{figure}

Consider the geodesic distance in G-space from Earth (point $E$) to Mars (point $M$) when Mars is near superior conjunction and the radar 
path $EM$ passes near the sun, as in Fig.~\ref{fig3}.  
In the absence of the 
   sun ($m = 0$), or when the sun is far from the radar 
   propagation path, 
   the G-space metric (\ref{iso3D}) is Euclidean 
   ($d\ell^{2}_{0}=dx^{2}+dy^{2}+dz^{2}$), and the geodesic 
   path from $E$ to $M$ is a straight line of some length 
   $\ell_{0}$.  With the sun in place near the propagation 
   path ($m=GM_{\odot}/c^{2}$), the G-space metric is given by 
   (\ref{iso3D}), and the geodesic distance from $E$ to $M$ is 
   increased to $\ell=\int_{E}^{M}d\ell$.  
   Because the speed of light is 
   everywhere $c$ for the single observer, the increase in path 
   length from $E$ to $M$ causes an additional radar delay $\delta 
   t= 2(\ell-\ell_{0})/c$ over and above what would be expected in 
   Euclidean space.  This is the excess delay 
   measured in the Shapiro relativistic radar echo experiment \cite{Shapiro}.

\begin{quote}

\textbf{Result IV:}\textit{\ In the single-observer picture, the 
   relativistic radar echo delay is attributed to an increase in 
   distance between the transmitter and target when a large mass (e.g.,the 
   sun) is brought near the radar propagation path.}

\end{quote}

\begin{figure}
   \includegraphics[width=3in]{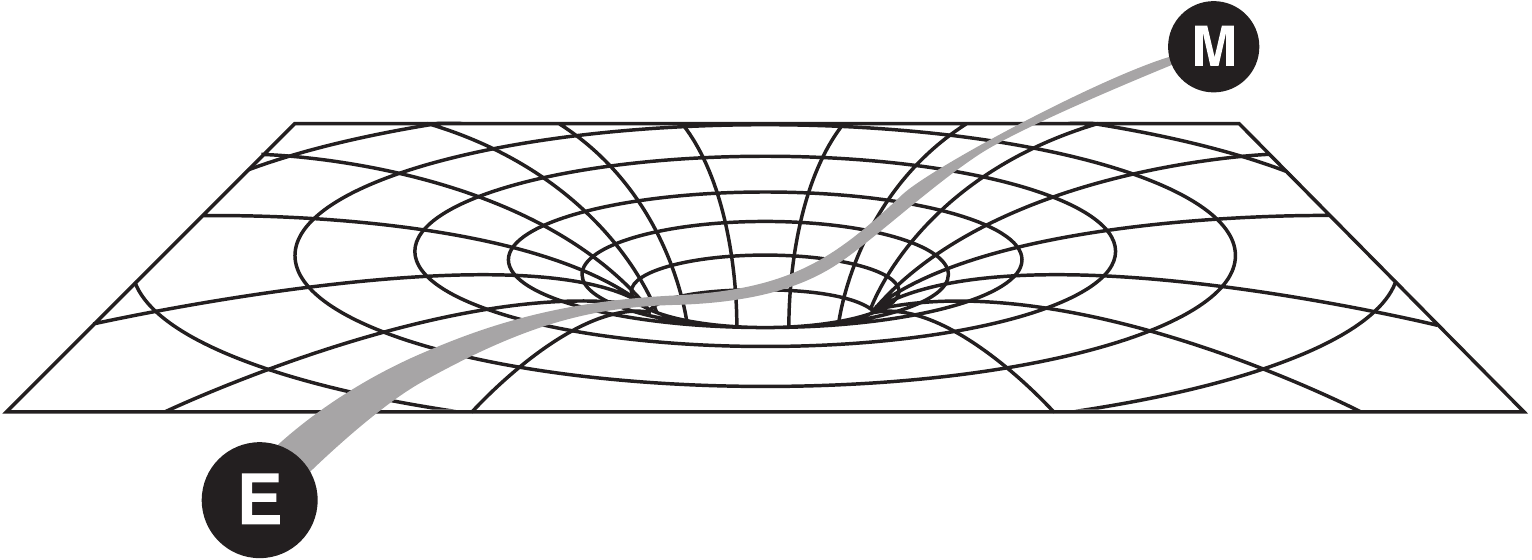}
   \caption{\label{fig4}Embedding diagram 
   for the $\theta=\pi/2$ plane of G-space in the neighborhood of the 
   sun (schematic and not to scale).  A light ray passing near the sun 
   is deflected because it travels a geodesic path on the embedding 
   surface.  A radar pulse is delayed because the sun increases the 
   geodesic distance from Earth $E$ to Mars $M$ when it is near the 
   radar propagation path.  The increased distance is attributed to 
   the depression of the embedding surface near the sun.}
 \end{figure}

   Both the bending of light and the excess radar echo delay are 
   nicely pictured by embedding the $\theta = \pi/2$ plane of 
   G-space in Euclidean three-space.  The embedding diagram is 
   sketched in Fig.~\ref{fig4}. 
The curvature of the surface causes deflection of 
   a light ray (a G-space geodesic) and the increased distance from 
   $E$ to $M$ is due to the depression of the surface near the mass 
   $M_{\odot}$.
   
   \begin{quote}

   \textbf{MO-Picture Interpretation}  Interpretation of the radar delay is more complicated 
   in MO-spacetime.  In MO-space, half of the delay is attributed 
   to an increased path length and half to 
   gravitational time dilation along the propagation path  
   \cite{Will2}.

   \end{quote}

   \section{\label{sec:ParticleMotion}PARTICLE MOTION IN G-SPACE}
   
   It is of interest to learn how material particles move in 
  the three-space for which light rays define geodesics.  We find 
  that the single-observer picture of particle motion is very much 
  closer in spirit to Newtonian mechanics than to the usual 
  formulation in general relativity.  This allows a clear comparison of 
  Newtonian and relativistic motions \emph{without approximation}.
   
  \subsection{Equations of motion}
   
  Particles travel on geodesics of MO-spacetime: 
   \begin{equation}
   	\delta\int_{}^{}d\bar{\tau} = 0 ,
   	\label{minproptime}
   \end{equation}
   where $\bar{\tau}$ is the proper time at the moving 
   particle as measured with a locally defined time standard (MO-picture). 
   Using equation (\ref{timescale}), 
 $d\bar{\tau}=\mathcal{R}d\tau$, and noting that the proper time in the 
 SO-picture ($d\tau^{2}=-ds^{2}/c^{2}$) obeys the same motional time 
 dilation formula as in special relativity,
 \begin{equation}
 d\tau = \sqrt{1-v^{2}/c^{2}}dt,
 \label{TimeDilation}
 \end{equation}
 where $v^{2}=d\ell^{2}/dt^{2}=g_{ij}(dx^{i}/dt)(dx^{j}/dt))$ is the squared speed of the particle in G-space, 
 equation (\ref{minproptime}) indicates that particle motion is 
 derivable from the three-space Lagrangian
\begin{eqnarray}
L &=&-\bar{m}c^{2}\mathcal{R}\left(1-\frac{v^{2}}{c^{2}}\right)^{1/2}
\nonumber \\ 
&=&-\bar{m}c^{2}\mathcal{R}\left(1-\frac{g_{ij}}
{c^{2}}\frac{dx^{i}}{dt}
\frac{dx^{j}}{dt} 
\right)^{1/2} ,
\label{lagrang}
\end{eqnarray}
here written in terms of G-space variables.
    We have added the inessential constant factor $-\bar{m}c^{2}$ for 
    later convenience of interpretation, where $\bar{m}$ is the rest mass in 
    the MO-picture (a constant).  From this Lagrangian we obtain 
    the momentum
\begin{eqnarray}
p_{k}&=&\frac{\partial L}{\partial (dx^{k}/dt)} \nonumber \\
&=& \frac{\bar{m}\mathcal{R}v_{k}}{\sqrt{1-v^{2}/c^{2}}}
\label{Momentum}
\end{eqnarray}
conjugate to coordinate $x^{k}$, where 
$v_{k}=g_{ki}v^{i}=g_{ki}dx^{i}/dt$. The Hamiltonian
\begin{eqnarray}
H&=& p_{k}v^{k}-L \nonumber \\
&=& \frac{\bar{m}c^{2}\mathcal{R}}{\sqrt{1-v^{2}/c^{2}}} \nonumber \\
&=& E,
\label{Ham}
\end{eqnarray}
is the conserved energy E of the particle. Note that in Gaussian space 
indices are lowered and raised with $g_{ij}$ and it's inverse $g^{ij}$, 
respectively. From the Lagrangian (\ref{lagrang}) there follows the 
equation of motion in G-space
     \begin{equation}
     	\frac{d^{2}x^{k}}{dt^{2}}+\Gamma^{k}_{\ ij}
     	\frac{dx^{i}}{dt}\frac{dx^{j}}{dt}
     	=-\left( \frac{\bar{m}c^{2}}{E}\right)^{2}
     	\frac{\partial \Phi}{\partial x_{k}} ,
     	\label{equmotion}
     \end{equation}
 where $\Gamma^{k}_{\ ij}=(1/2)g^{kl}(\partial_{j}g_{li}+
 \partial_{i}g_{lj}-\partial_{l}g_{ij})$ are the G-space 
Christoffel symbols, we have used the constancy of $E$ in the 
 derivation, and 
 \begin{equation}
 \Phi = \frac{1}{2}c^{2}\mathcal{R}^{2}
 \label{Newtonian}
 \end{equation}
 is the ``Newtonian gravitational potential'' in G-space (the 
 justification for this terminology will become apparent in Sec.~\ref{sec:GravitationalPotential}).  From Eq.~(\ref{equmotion}) we see that, as the particle 
 approaches light speed ($E >> \bar{m}c^{2}$), the term on the 
 right vanishes and the particle moves on a geodesic of G-space, as do photons.
 
 Notice that the Lagrangian
 \begin{equation} 
 L=-mc^{2}\sqrt{1-v^{2}/c^{2}}  ,
 \label{Lagr}
 \end{equation} 
 the energy
\begin{equation}
E=\frac{mc^{2}}{\sqrt{1-v^{2}/c^{2}}}=\sqrt{c^{2}p^{2}+m^{2}c^{4}}  , 
\label{Ener}
\end{equation}
and the momentum
\begin{equation} 
\mathbf{p}=\frac{m\mathbf{v}}{\sqrt{1-v^{2}/c^{2}}}
\label{Mo}
\end{equation}
in G-space all have the 
same forms as in special relativity, provided we identify the 
position-dependent quantity
\begin{equation} 
m(\mathbf{x})=\bar{m}\mathcal{R}(\mathbf{x})
\label{RestMass}
\end{equation} 
as the \emph{rest mass of the particle in the SO-picture}.  On this interpretation, the 
position-dependent rest energy of the particle, 
$mc^{2}=\bar{m}c^{2}\mathcal{R}$, plays 
the role of a gravitational potential energy in G-space, as can be 
seen in the non-relativistic ($v^{2} \ll c^{2}$) limits of the 
lagrangian (\ref{Lagr}) [$L=mv^{2}/2-m(\mathbf{x})c^{2}$] and the energy (\ref{Ener})
[$E=mv^{2}/2 +m(\mathbf{x})c^{2}$]. 

In fact, the interpretation of the 
rest energy $m(\mathbf{x})c^{2}$ as a gravitational potential energy 
is fully relativistic. The G-space energy $E$ and three-momentum $p^{i}$ 
combine to form a four-vector momentum
\begin{equation}
P^{\mu}=(\frac{E}{c}, p^{i}) = m\frac{dx^{\mu}}{d\tau} ,
\label{FourMomentum}
\end{equation}
in the SO-picture. This is related to the MO-picture momentum $\bar{P}^{\mu}\equiv 
\bar{m}dx^{\mu}/d\bar{\tau}$ by the conformal transformation law
\begin{equation}
P^{\mu}=\mathcal{R}^{2}\bar{P}^{\mu} .
\label{MomentumTrans}
\end{equation}
In the absence of non-gravitational forces, 
the four-momentum $\bar{P}^{\mu}$ is parallel transported in 
MO-spacetime [$ d\bar{P}^{\mu}/d\bar{\tau}+\bar{\Gamma}^{\mu}_{\ \alpha 
\beta}\bar{P}^{\alpha}dx^{\beta}/d\bar{\tau}=0$]. But this equation 
is \emph{not} conformally invariant.  In the SO-picture, the 
four-momentum (\ref{FourMomentum}) obeys the equation of motion
\begin{equation}
\frac{DP^{\mu}}{d\tau} \equiv \frac{dP^{\mu}}{d\tau}+\Gamma^{\mu}_{\ \alpha \beta}P^{\alpha}
\frac{dx^{\beta}}{d\tau}=-\frac{\partial (mc^{2})}{\partial x_{\mu}} 
\label{EOMmomentum}
\end{equation}
in which the rest energy $m(\mathbf{x})c^{2}$ manifestly plays the 
role of a scalar potential energy.

\begin{quote}

\textbf{Result V:}  \textit{Although there is no true gravitational 
force in the MO-picture, i.e., the momentum $\bar{P}^{\mu}$ obeys the ``law of inertia'' $D\bar{P}^{\mu}/d\bar{\tau}=0$ and is constant in a local inertial frame, the momentum $P^{\mu}=\mathcal{R}^{2}\bar{P}^{\mu}$ of the SO-picture changes due to the action of a gravitational four-force $f^{\mu}=-\partial V/\partial x_{\mu}$ derivable from a scalar potential $V=m(\mathbf{x})c^{2}$ that is the rest energy of the particle in this picture:}
\begin{equation}
\frac{DP^{\mu}}{d\tau}\equiv\frac{dP^{\mu}}{d\tau}+
\Gamma^{\mu}_{\ \alpha \beta}P^{\alpha}
\frac{dx^{\beta}}{d\tau}=-\frac{\partial V}{\partial x_{\mu}} .
\label{EOMmomentum}
\end{equation}

\end{quote}

Notice that the ``Newtonian gravitational potential'' 
$\Phi$, the ``gravitational potential energy'' $V$, and the ``atomic 
rest-clock rate'' $\mathcal{R}$ all contain the same information: 
\begin{equation}
\frac{2\Phi}{c^{2}}=\left(\frac{V}{\bar{m}c^{2}}\right)=\mathcal{R}^{2} .
\label{Relation}
\end{equation} 
The potential energy $V(\mathbf{x})$ determines the rate of change of momentum $P^{\mu}$, the gravitational potential $\Phi$ (together with the particle mass $\bar{m}$ and its energy $E$) determine the three-acceleration of the particle in G-space [Eq.~(\ref{equmotion})], and the scale factor $\mathcal{R}(\mathbf{x})$ is the ticking rate of a free-running atomic clock on the time scale of observer O's local clock.

\subsection{Newtonian Analogy}
  
  The exact relativistic equation of motion in G-space 
  (\ref{equmotion}) is quite similar to the Newtonian equation of 
  motion in curvilinear coordinates
  \begin{equation}
  	\frac{d^{2}x^{k}}{dt^{2}}
  	+\Gamma^{k}_{ij}
  	\frac{dx^{i}}{dt}\frac{dx^{j}}{dt}
   	= -\frac{\partial \Phi}{\partial x_{k}} .
  	\label{Newteqn}
  \end{equation}
Formally, the only difference is the constant 
factor $(\bar{m}c^{2}/E)^{2}$ on the right in the G-space equation 
(\ref{equmotion}).  When the particle moves slowly ($E \approx 
\bar{m}c^{2}$), the two equations are formally identical, and when the 
particle moves fast ($E \gg \bar{m}c^{2}$), the factor 
$(\bar{m}c^{2}/E)^{2}$ turns off   the gravitational acceleration.  The quantity on the left in 
equation (\ref{equmotion}) is the acceleration of the particle in 
G-space (the absolute acceleration), which, in this case, is the 
acceleration due to gravity $g^{k}$ in G-space. Thus we have

\begin{quote}

\textbf{Result VI:}  \textit{For a slowly moving particle ($E \approx 
\bar{m}c^{2}$), the acceleration due to gravity in Gaussian space is 
derivable from the ``Newtonian gravitational potential'' 
$\Phi=c^{2}\mathcal{R}^{2}/2$ as}
\begin{equation}
\mathbf{g}_{0}=-\nabla \Phi .
\label{RestGravAccel}
\end{equation}
\textit{As the speed of the particle increases, the gravitational 
acceleration $\mathbf{g}$ decreases by the factor 
$(\bar{m}c^{2}/E)^{2}$ which is a constant of the motion:}
\begin{equation}
\mathbf{g} = \left(\frac{\bar{m}c^{2}}{E}\right)^{2}\mathbf{g}_{0} .
\label{GravAccel}
\end{equation}
\textit{As the particle approaches the speed of light ($E\gg 
\bar{m}c^{2}$), the gravitational acceleration tends to zero and the 
trajectory of the particle approaches a G-space geodesic.}

\end{quote}

\begin{figure}
\includegraphics[width=3in]{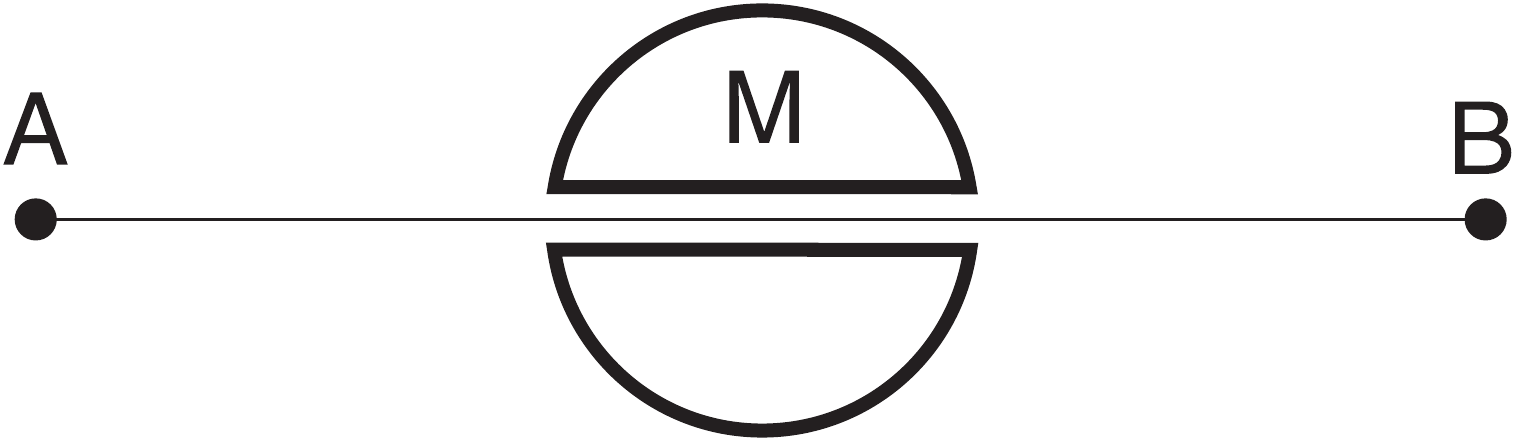}
\caption{\label{fig5}The straight path from $A$ to $B$ passes through 
mass $M$.  With the mass in place, light takes longer to propagate 
from $A$ to $B$ than without the mass present, whereas a 
non-relativistic material particle travels from $A$ to $B$ faster 
with $M$ present than without.  How can light be slowed and particles 
hastened along the same path in the same gravitational field?}
\end{figure}

Result VI helps to clarify 
  certain curious results of general relativity.  
  Consider the problem depicted in Fig.~\ref{fig5}, which is said to 
  have puzzled Einstein some years after the development of 
  general relativity. 

Light propagating from $A$ to $B$ through the 
  mass $M$ has its propagation time delayed from what it would be 
  without the mass present, but a non-relativistic particle traveling 
  the same path arrives at $B$ sooner than it would without the 
  mass $M$ present. \emph{How can light be slowed and particles be hastened
  along the same path in the same gravitational field?}  In the 
  SO-picture the answer is clear.  The delay for 
  light, as for radar, is 
  due to the increased G-space distance between $A$ and $B$ when the 
  mass $M$ is in place.  The material particle must also travel the 
  longer path between $A$ and $B$, but unlike the photon which 
  experiences no gravitational acceleration in G-space ($\mathbf{g}=0$), 
  a slow material 
  particle  experiences the Newtonian 
 acceleration $\mathbf{g}_{0}=-\nabla \Phi$, which increases the particle's 
 speed as it nears and 
  crosses the mass $M$, thus decreasing the travel time.  Hence the 
  particle arrives sooner and the photon later because the particle 
  experiences a gravitational acceleration in G-space and the photon 
  does not.
  
 \begin{quote}

 \textbf{MO-Picture Interpretation:}  For comparison, we note that the acceleration due to gravity in 
 MO-space (the absolute acceleration on the time scale of the 
 coordinate time $\bar{t}$),
 \begin{eqnarray}
 \bar{g}^{k}&=& \frac{d^{2}x^{k}}{d\bar{t}^{2}}+\bar{\Gamma}^{k}_{ij}\frac{dx^{i}}{d\bar{t}}
 \frac{dx^{j}}{d\bar{t}} \nonumber \\ 
  &=& -\frac{\partial \bar{\phi}}{\partial x_{k}}+\frac{1}{\bar{\phi}}
 \left(\frac{\partial \bar{\phi}}{\partial x^{i}}
 \frac{dx^{i}}{d\bar{t}}\right)\frac{dx^{k}}{d\bar{t}} ,
 \label{MOequMotion}
 \end{eqnarray}
 is not the gradient of a scalar potential alone, but in addition 
 contains a velocity-dependent term that enforces the speed limit 
 $d\bar{\ell}/d\bar{\tau}_{0}\le c$ in MO-space [here 
 $\bar{\Gamma}^{k}_{ij}=(1/2)\bar{g}^{kl}(\partial_{j}\bar{g}_{li}+
 \partial_{i}\bar{g}_{lj}-\partial_{l}\bar{g}_{ij})$ are the 
Christoffel symbols  in MO-space and
 \begin{equation}
 \bar{\phi}=-\frac{1}{2}c^{2}\bar{g}_{00}
 \label{MOpotential}
 \end{equation}
 is the low-speed gravitational potential in this space].

 \end{quote}
 
 There are two fundamental differences between the G-space equation 
 of motion (\ref{equmotion}) and the Newtonian equation of motion 
 (\ref{Newteqn}).  The first (a local difference) is the factor 
 $(\bar{m}c^{2}/E)^{2}$ discussed above.  The second difference (a 
 global one) is the curvature of Gaussian space for equation 
 (\ref{equmotion}) whereas the space for the Newtonian equation 
 (\ref{Newteqn}) is flat.  In the following section we learn that 
 this difference accounts for the perihelion precession of planetary 
 orbits.

  \subsection{Relativistic Kepler Problem}
  
  In Schwarzschild geometry (\ref{eq.Scharz}), the ``Newtonian 
  gravitational potential'' (\ref{Newtonian}) for an observer at infinity is
  \begin{equation}
  \Phi=-\frac{GM}{r}+\frac{c^{2}}{2}	 .
  	\label{schwzpot}
  \end{equation}
 Apart from the additive constant $c^{2}/2$, 
 this is formally identical to the Newtonian gravitational potential 
 for this problem [the additive constant is, of course, arbitrary in 
 Newtonian theory, but fixed by equations (\ref{Rate}) and 
 (\ref{Newtonian}) in 
 the SO-picture of general relativity].  The gravitational acceleration in G-space, 
 Eq.~(\ref{GravAccel}), reads
 \begin{equation}
 \mathbf{g}=\left(\frac{\bar{m}c^{2}}{E}\right)^{2}\nabla 
 \left(\frac{GM}{r}\right) .
 \label{GravAccelII}
 \end{equation}
 The constant factor $(\bar{m}c^{2}/E)^{2}$ in this equation is equivalent to a 
 change of the central mass $M$ by this factor; a change which is entirely 
 negligible for planetary motions in the solar system (the change is 
 less than one part in $10^{7}$ for planet Mercury in the sun's 
 field, and $M_{\odot}$ is known only to about four significant 
 figures anyway).  Moreover, a small change in the strength of the 
 Newtonian inverse-square acceleration makes no contribution to the 
 perihelion motion (such a change acting alone still gives closed 
 elliptical orbits).  Therefore, the precession of perihelion can only 
 be attributed to the non-Euclidean character of G-space:
 
 \begin{quote}

  \textbf{Result VII:} \textit{In the single-observer picture, the 
 relativistic precession of perihelion of planetary orbits is due 
 exclusively to the curvature of Gaussian space.}

 \end{quote}
 
    Let us show this explicitly.  In the $\theta = \pi/2$ plane of 
  Schwarzschild space, we write the G-space metric, 
  Eq.(\ref{3Dmetric1}), as
  \begin{equation}
  		d\ell^{2} = \frac{1}{(1-2m^{*}/r)} \left[
	\frac{dr^{2}}{(1-2m^{*}/r)} + r^{2}d\phi^{2} \right] .
  	\label{mmetric}
  \end{equation}
  with a star on $m^{*}$ so that we can ``turn off'' the curvature of 
  G-space by setting $m^{*} = 0$, without changing the gravitational 
  potential (\ref{schwzpot}) [this is, of course, an artificial 
  procedure, but one that allows us to isolate the effect of G-space 
  curvature].
  
 The orbit equation for the variable $u(\phi) = 1/r(\phi)$ is derived 
 from from the equation of motion (\ref{equmotion}) in the usual way.  The result is
   \begin{equation}
  	\frac{d^{2} u}{d\phi^{2}} + u = \frac{GM}{h^{2}} + 
  	3m^{*} u^{2},
  	\label{orbpara}
  \end{equation}
  where
  \begin{equation}
  	h = \frac{r^{3}}{r-2m^{*}} \left( \frac{d \phi}{dt}
  	\right)
  	\approx r^{2}\frac{d\phi}{dt}
  	  	\label{const} 
  \end{equation}
  is a constant of the motion (the angular momentum  per unit mass), and we have replaced $(\bar{m}c^{2}/E)^{2}$ 
  by unity because the difference is negligible for any planetary 
  motion in the solar system.
  
  The first term on the right in (\ref{orbpara}) (the Newtonian term) 
  comes from the potential  (\ref{schwzpot}), whereas the second 
  term (the relativistic term) comes from the G-space metric (\ref{mmetric}) 
  and gives rise to the observed advance of 
  perihelion.  But, if we turn off the curvature of G-space by 
  setting $m^{*} = 0$, equation (\ref{orbpara}) is then the 
  Newtonian orbit equation with no precession of perihelion.  In this 
  sense the precession [$\delta\phi = 6\pi 
  m^{*}/a(1-\epsilon^{2})$ per revolution, where $a$ is the 
  semi-major axis and $\epsilon$ the eccentricity of the orbit] is a 
  direct measure of the curvature of G-space.
  
\begin{quote}

  \textbf{MO-Picture Interpretation:}  Interpretation of perihelion precession is more complicated in 
 MO-spacetime.  Will \cite{Will3} identifies three separate 
  significant contributions to the perihelion precession (three 
  significant terms in the PPN expansion): (1) curvature of MO-space, 
  (2) a velocity dependent part of the gravitational force, and (3) a 
  non-linear  term proportional to the square of the gravitational 
  potential; the relative contribution of these effects being 
  coordinate dependent.   

\end{quote}

 It is noteworthy that the single-observer 
  interpretation attributes the perihelion advance to a single 
  cause (the curvature of G-space) and this interpretation is exact
  rather than approximate.

\section{\label{sec:GravitationalPotential}THE GRAVITATIONAL POTENTIAL IN GAUSSIAN SPACE}

Let us derive the field equation for the ``Newtonian gravitational 
potential''
\begin{equation}
\Phi=\frac{1}{2}c^{2}\mathcal{R}^{2}
\label{NewtPotential}
\end{equation}
in G-space.  The exact result is surprisingly simple and familiar.

To begin, the metric $d\bar{s}^{2}$ in the MO-picture is written in 
terms of G-space variables as
\begin{equation}
d\bar{s}^{2}=\mathcal{R}^{2}ds^{2}=\mathcal{R}^{2}\left(-c^{2}dt^{2}
+g_{ij}dx^{i}dx^{j}\right)  .
\label{MOmetrictwo}
\end{equation}
Hence the metric tensor in the MO-picture reads
\begin{eqnarray}
\bar{g}_{00}&=& -\mathcal{R}^{2} , \nonumber \\
\bar{g}_{ij}&=& \mathcal{R}^{2}g_{ij} , \nonumber \\
\bar{g}_{0i}&=& \bar{g}_{i0} = 0 .
\label{MetricTensor}
\end{eqnarray}
From this we construct the 00-component of the Ricci tensor in the 
MO-picture.  The result,
\begin{equation}
\bar{R}_{00}=\frac{1}{2\mathcal{R}^{2}}\left[\frac{1}{\sqrt{g}}
\frac{\partial}{\partial x^{k}}\left(\sqrt{g}g^{kl}
\frac{\partial \mathcal{R}^{2}}{\partial x^{l}}\right)\right]
=\frac{\nabla^{2}\Phi}{c^{2}\mathcal{R}^{2}} ,
\label{R00}
\end{equation}
is the left hand side of the Einstein equation 
\begin{equation}
\bar{R}_{00}=8\pi G(\bar{T}_{00}-\bar{g}_{00}\bar{T}/2)/c^{4}. 
\label{00Einstein}
\end{equation}
 To evaluate the right hand side of this equation in the SO-picture we 
need the conformal transformation law for the stress-energy tensor 
$\bar{T}_{\alpha\beta}$.  The simple case of pressureless fluid
 ($\bar{T}^{\mu\nu}=\bar{\rho}_{0}\bar{U}^{\mu}\bar{U}^{\nu}$) 
 determines this transformation law as follows.  The transformation for 
 the mass density in the rest frame of the fluid 
 [which is $\bar{\rho}_{0}=d\bar{m}/(d\bar{\ell})^{3}$ in the MO-picture and 
$\rho_{0}=dm/(d\ell)^{3}$ in the SO-picture] is determined by the 
transformation laws for mass ($dm=\mathcal{R}d\bar{m}$) and length
 ($d\ell=d\bar{\ell}/\mathcal{R}$):
 \begin{equation}
 \rho_{0}=\mathcal{R}^{4}\bar{\rho}_{0} .
 \label{RhoTrans}
 \end{equation}
 The transformation for the fluid four-velocity $\bar{U}^{\mu}$ 
 [which is $\bar{U}^{\mu}=dx^{\mu}/d\bar{\tau}$ in the MO-picture and
$U^{\mu}=dx^{\mu}/d\tau$ in the SO-picture] is determined by the 
transformation law for proper time ($d\bar{\tau}=\mathcal{R}d\tau$):
\begin{equation}
U^{\mu}=\mathcal{R}\bar{U}^{\mu} .
\label{FourTrans}
\end{equation}
Hence the stress-energy tensor in the SO-picture, $T^{\mu\nu}\equiv 
\rho_{0}U^{\mu}U^{\nu}$, is related to the stress-energy tensor $\bar{T}^{\mu\nu}=\bar{\rho}_{0}\bar{U}^{\mu}\bar{U}^{\nu}$ of the MO-picture as 
\begin{equation}
T^{\mu\nu}=\mathcal{R}^{6}\bar{T}^{\mu\nu} ,
\label{StressTrans}
\end{equation}
and lowering indices with the metric, which transforms as 
$g_{\alpha\beta}=\bar{g}_{\alpha\beta}/\mathcal{R}^{2}$, gives
$T_{\mu\nu}=\mathcal{R}^{2}\bar{T}_{\mu\nu}$ and $T=T^{\mu}_{\mu}=\mathcal{R}^{4}\bar{T}$.  All of these conformal transformation rules are consistent with the general transformation rule derived in the Appendix.  Finally, the 00-component of Einstein's equation (\ref{00Einstein}) becomes the field 
equation for the gravitational potential $\Phi$ in Gaussian space:

\begin{quote}

\textbf{Result VIII:} \textit{The acceleration due to gravity for a 
particle at rest in Gaussian space, $\mathbf{g}_{0}=-\nabla\Phi$, is 
derivable from a potential $\Phi$ which satisfies the same linear 
Poisson equation as in Newtonian gravitation theory,}
\begin{equation}
\nabla^{2}\Phi =4\pi G\rho_{g} ,
\label{Poisson}
\end{equation}
\textit{with active gravitational mass density}
\begin{equation}
\rho_{g}=(2T_{00}-g_{00}T)/c^{2}
\label{GravMass}
\end{equation}
\textit{acting as the source of gravitational field.  This is the 
ultimate justification for calling $\Phi$ the ``Newtonian 
gravitational potential''.}

\end{quote}

In view of this result, the equations for the ``Newtonian gravitational field'' 
$\mathbf{g}_{0}$ in G-space can be written as
\begin{eqnarray}
\nabla\cdot\mathbf{g}_{0}&=& -4\pi G\rho_{g} , \nonumber \\
\nabla \times \mathbf{g}_{0} &=& 0 .
\label{GravFieldEQUS}
\end{eqnarray}
The first of these, when integrated over a volume $V$ with closed 
surface $S$, gives the gravitational Gauss law in G-space:

\begin{quote}

\textbf{Result IX:} \textit{The flux of $\mathbf{g}_{0}$ through any 
closed surface $S$ in G-space equals $4\pi G$ times the total active 
gravitational mass $M_{g}$ inside the surface:}
\begin{equation}
\oint_{S} \mathbf{g}_{0}\cdot d\mathbf{a} = 4\pi GM_{g} ,
\label{GaussLaw}
\end{equation}
\textit{where}
\begin{equation}
M_{g}=\int_{V} \rho_{g}\sqrt{g}d^{3}x  .
\label{GravMass}
\end{equation}

\end{quote}

This result follows because the divergence theorem holds true in a 
non-Euclidean three-space as well as a Euclidean one.

Result IX may be used to operationally define the mass $M_{g}$ inside a 
closed surface in G-space in terms of the field $\mathbf{g}_{0}$ on 
that surface (without the space being asymptotically flat), or to 
calculate the gravitational field $\mathbf{g}_{0}$ in cases of simple 
symmetry, just as the electric field is calculated using the 
electrostatic Gauss law in a flat space.

\begin{quote}

\textbf{MO-Picture Interpretation}  For comparison, we note that the \emph{exact} field equation for the 
``gravitational potential'' $\bar{\phi}=-c^{2}\bar{g}_{00}/2$ in 
MO-space (the potential whose gradient $\bar{g}_{k}=-\partial 
\bar{\phi}/\partial x^{k}$ is the rest acceleration due to gravity in 
this space) is the 00-component of Einstein's equation 
$\bar{R}_{\mu\nu}=(8\pi 
G/c^{2})(\bar{T}_{\mu\nu}-\bar{g}_{\mu\nu}\bar{T}/2)$, namely the 
non-linear equation
\begin{equation}
\bar{\nabla}^{2}\bar{\phi}-
\frac{\bar{\nabla}\bar{\phi}\cdot \bar{\nabla}\bar{\phi}}{2\bar{\phi}}
=4\pi G\bar{\rho}_{g} ,
\label{MOPoisson}
\end{equation}
where $\bar{\rho}_{g}= (2\bar{T}_{00}-\bar{g}_{00}\bar{T})/c^{2}$ is 
the density of active gravitational mass in MO-space.  There do not 
appear to be any simple exact results in the MO-picture analogous to Results VIII and IX in 
SO-space.
\end{quote}

The linear Poisson equation (\ref{Poisson}) for the gravitational potential $\Phi$ in G-space is an exact, fully relativistic result valid for any static gravitational field.  This result is surprising because the Einstein equation, from which it is derived, is clearly non-linear.

 \section{\label{sec:Electrodynamics}ELECTRODYNAMICS}
 
 In MO-spacetime the electromagnetic field $\bar{F}_{\mu \nu}$ is governed 
 by Maxwell's equations
 \begin{subequations}
 \label{MOMaxwells}
 \begin{equation}
 \bar{\nabla}_{\omega}\bar{F}_{\mu \nu}+
 \bar{\nabla}_{\nu}\bar{F}_{\omega\mu}+
 \bar{\nabla}_{\mu}\bar{F}_{\nu \omega} = 0,
 \label{MOmax1}
 \end{equation}
 and
 \begin{equation}
 \bar{\nabla}_{\nu}\bar{F}^{\mu\nu}=-\frac{4\pi}{c}\bar{J}^{\mu},
 \label{MOmax2}
 \end{equation}
 \end{subequations}
 where $\bar{J}^{\mu}=(c\bar{\rho}, \bar{\mathbf{j}})/\sqrt{-\bar{g}_{00}}$ 
 is the four-current density, determined by the charge density 
 $\bar{\rho}$ and current density $\bar{\mathbf{j}}$ in MO-space and 
 $\bar{\nabla}_{\omega}$ is the covariant derivative based on the MO-metric $\bar{g}_{\alpha\beta}$.
The motion of a particle of charge $\bar{q}$ and mass $\bar{m}$ is 
 described by the Lorentz equation of motion
 \begin{equation}
 \frac{d\bar{P}^{\mu}}{d\bar{\tau}}+\bar{\Gamma}^{\mu}_{\ \alpha\beta}
 \bar{P}^{\alpha}\frac{dx^{\beta}}{d\bar{\tau}}=
 \frac{\bar{q}}{c}{{\bar{F}}^{\mu}}_{ \ \nu}\frac{dx^{\nu}}{d\bar{\tau}},
 \label{MOeom}
 \end{equation}
where $\bar{P}^{\mu}= \bar{m}dx^{\mu}/d\bar{\tau}$ is the four-momentum 
of the particle and $\bar{\tau}$ it's proper time.

The Maxwell equations (\ref{MOMaxwells}) are 
\emph{conformally invariant}.  By this is meant that, if under the 
conformal transformation 
$g_{\alpha\beta}=\bar{g}_{\alpha\beta}/\mathcal{R}^{2}$ the field 
tensor and current density transform as
\begin{equation}
F_{\mu\nu}=\bar{F}_{\mu\nu}
\label{FT}
\end{equation}
and 
\begin{equation}
J^{\mu}=\mathcal{R}^{4}\bar{J}^{\mu},
\label{CD}
\end{equation}
respectively [or equivalently 
${F^{\mu}}_{\nu}=\mathcal{R}^{2}{{\bar{F}}^{\mu}}_{ \nu}$ 
(or $F^{\mu\nu}=\mathcal{R}^{4}\bar{F}^{\mu\nu}$) and 
$J_{\mu}=\mathcal{R}^{2}\bar{J}_{\mu}$], then, after the conformal 
transformation, Maxwell's equations have the same four-space forms as 
before, namely
\begin{subequations}
\label{SOMaxwells}
\begin{equation}
 \nabla_{\omega}F_{\mu 
 \nu}+\nabla_{\nu}F_{\omega\mu}+
 \nabla_{\mu}F_{\nu \omega} = 0,
 \label{SOmax1}
 \end{equation}
 and
 \begin{equation}
 \nabla_{\nu}F^{\mu\nu}=-\frac{4\pi}{c}J^{\mu},
 \label{SOmax2}
 \end{equation}
 \end{subequations}
where $\nabla_{\nu}$ is the covariant derivative appropriate to 
the SO-metric $g_{\alpha\beta}$ and $J^{\mu}=(c\rho, \mathbf{j})$ is 
the four-current density in SO-spacetime.

The equation of motion (\ref{MOeom}) is \emph{not} conformally 
invariant.  In SO-spacetime the equation of motion reads
\begin{equation}
\frac{dP^{\mu}}{d\tau}+\Gamma^{\mu}_{\ \alpha\beta}
P^{\alpha}\frac{dx^{\beta}}{d\tau}=-\frac{\partial V}{\partial 
x_{\mu}}+\frac{q}{c}{F^{\mu}}_{\nu}\frac{dx^{\nu}}{d\tau} ,
\label{SOeom}
\end{equation}
where $V(\mathbf{x})=m(\mathbf{x})c^{2}$ is the rest energy of the particle in 
this picture, and we have used ${F^{\mu}}_{\nu}=
\mathcal{R}^{2}{{\bar{F}}^{\mu}}_{\nu}$, $d\tau=d\bar{\tau}/\mathcal{R}$,
  $P^{\mu}=\mathcal{R}^{2}\bar{P}^{\mu}$, and $q=\bar{q}$ in the derivation.
  The charges $q$ and $\bar{q}$ in the SO- and MO-pictures, respectively, are set equal to one another  (as opposed to some other transformation law $q=\mathcal{R}^{s}\bar{q}$ with 
  $s\ne 0$) so that the mere motion of a single charge does not 
  violate charge conservation in the SO-picture.  This also follows from the charge conservation law $\nabla_{\mu}J^{\mu}=0$ in SO-spacetime, which is derivable from (\ref{SOmax2}).

This completes the four-space formulation of classical electrodynamics 
in the SO-picture.  But our interest here centers on the forms taken 
by Maxwell's equations in the Gaussian three-space and how these compare 
with the corresponding equations in MO-space.  As we shall see, the 
conformal invariance of the 4-space Maxwell equations does not 
imply that the 3-space Maxwell equations are the same in the MO- and 
SO-pictures because the metric elements $\bar{g}_{00}$ and 
$g_{00}$ ($=-1$) are different in the two pictures.

\subsection{Maxwell's Equations in MO-Space}

There have been several three-space (or ``3+1'') formulations of Maxwell's 
equations in the MO-picture, all of which seem to have certain 
``unphysical'' features.  Here we consider two of them.

Perhaps the earliest formulation is the one in which the 
electrodynamic equations in a static gravitational field are formally 
identical to the Maxwell equations in a material medium:
\begin{subequations}
\label{MO3Maxwells}
\begin{eqnarray}
\bar{\nabla}\cdot\tilde{\mathbf{D}}&=&4\pi \tilde{\rho} , 
\label{MatMax1} \\
\bar{\nabla}\cdot\tilde{\mathbf{B}}&=&0 ,
\label{MatMax2} \\
\bar{\nabla}\times\tilde{\mathbf{H}}-\frac{1}{c}
\frac{\partial \tilde{\mathbf{D}}}{\partial \bar{t}}
&=&\frac{4\pi}{c}\tilde{\mathbf{j}} ,
\label{MatMax3} \\
\bar{\nabla}\times\tilde{\mathbf{E}}+\frac{1}{c}
\frac{\partial \tilde{\mathbf{B}}}{\partial \bar{t}}&=&0  ,
\label{MatMax4}
\end{eqnarray}
\end{subequations}
with constitutive relations
\begin{eqnarray}
\tilde{\mathbf{D}}&=&\frac{\tilde{\mathbf{E}}}{\sqrt{-\bar{g}_{00}}}
=\tilde{\epsilon}\tilde{\mathbf{E}}  , 
\label{Dconst} \\
\tilde{\mathbf{H}}&=&\frac{\tilde{\mathbf{B}}}{\sqrt{-\bar{g}_{00}}}
=\frac{\tilde{\mathbf{B}}}{\tilde{\mu}}  ,
\label{Hconst}
\end{eqnarray}
corresponding to a medium with ``dielectric constant'' 
$\tilde{\epsilon}$ and ``magnetic permeability'' $\tilde{\mu}$ given by
\begin{equation}
\tilde{\epsilon}=\frac{1}{\tilde{\mu}}=\frac{1}{\sqrt{-\bar{g}_{00}}} .
\label{Constitutive}
\end{equation}
In equations (\ref{MO3Maxwells}), 
$\bar{\nabla}\cdot\tilde{\mathbf{D}}$ and 
$\bar{\nabla}\times\tilde{\mathbf{E}}$ are the divergence and curl in 
MO-space (the three-space with metric $\bar{g}_{ij}$).  Equations 
(\ref{MO3Maxwells}) are derived and studied in Landau and 
Lifshitz \cite{Landau}.  We shall refer to them as the ``material 
Maxwell equations''.

It is pleasant that the Maxwell equations (\ref{MO3Maxwells}) 
take the familiar forms for electrodynamics in a material medium, but 
the interpretations of $\tilde{\epsilon}$ as a dielectric constant and 
of $\tilde{\mu}$ as a magnetic permeability of space are without 
physical foundation (empty space contains neither the electric charges 
to be a dielectric medium nor the electric currents to be a magnetic 
one).  Moreover, the electric field $\tilde{\mathbf{E}}$ in the 
material Maxwell equations is not the electric field measured locally 
(it is more a formal field than a physical one).  For 
these reasons, the material Maxwell equations must be viewed as 
``unphysical''.

An important step toward physical clarity was taken by Thorne and 
Macdonald \cite{Thorne1} and others \cite{Thorne2}, in connection with 
the membrane paradigm for black holes. In this approach, the electric and magnetic fields
\begin{equation}
\bar{E}^{k}=\frac{{\bar{F}}^{k}_{ 0}}{\sqrt{-\bar{g}_{00}}}
=\frac{\tilde{E}^{k}}{\sqrt{-\bar{g}_{00}}}
\label{LocalE}
\end{equation}
and
\begin{equation}
\bar{B}^{k}=\frac{1}{2}\bar{e}^{k}_{ij}\bar{F}^{ij}=\tilde{B}^{k},
\label{LocalB}
\end{equation}
are those measured locally using the traditional time and length 
standards of the MO-picture.  For these fields, the 
three-space Maxwell equations read
\begin{subequations}
\label{MOKipMaxwells}
\begin{eqnarray}
\bar{\nabla}\cdot\bar{\mathbf{E}}&=&4\pi\bar{\rho}  ,
\label{LocalMax1} \\
\bar{\nabla}\cdot\bar{\mathbf{B}}&=&0  ,
\label{LocalMax2} 
\end{eqnarray}
\begin{eqnarray}
\frac{1}{c}\frac{\partial \bar{\mathbf{B}}}{\partial \bar{t}}
&=&-\bar{\nabla}\times(\alpha\bar{\mathbf{E}})  ,
\label{LocalMax3} \\
\frac{1}{c}\frac{\partial \bar{\mathbf{E}}}{\partial \bar{t}}
&=&\bar{\nabla}\times (\alpha\bar{\mathbf{B}})-\frac{4\pi}{c}\alpha\bar{\mathbf{j}}  ,
\label{LocalMax4}
\end{eqnarray}
\end{subequations}
where $\alpha=\sqrt{-\bar{g}_{00}}$ is called the ``lapse 
function,'' and $\bar{\rho}$ ($= \tilde{\rho}$) and 
$\bar{\mathbf{j}}$ ($=\tilde{\mathbf{j}}/\sqrt{-\bar{g}_{00}}$) are 
the locally measured charge and current densities.

Equations (\ref{MOKipMaxwells}) are not in the form of 
Maxwell's equations for a material medium. However, they do ascribe 
electric and magnetic properties to empty space, e.g., the 
Schwarzschild surface behaves in many respects like a conducting 
membrane, and this is often useful, though in reality there are no 
electric currents in this surface.

It is desirable, for aesthetic as well as practical reasons, to have a 
three-space formulation of Maxwell's equations that is ``true to the 
physics'' in the sense that it does not assign electric or magnetic 
properties to empty space where there are no charges or currents.  In 
the next section we show that the SO-picture provides such a 
formulation.

\subsection{Maxwell's Equations in Gaussian Space}

In order to define electric and magnetic fields appropriate to the 
SO-picture, we multiply equation of motion (\ref{SOeom}) by 
$d\tau/dt=mc^{2}/E$ and note that, for the SO-metric (\ref{SOtrans}), 
only the spatial components $\Gamma^{k}_{\ ij}$ of the Christoffel symbols 
$\Gamma^{\mu}_{\ \alpha\beta}$ are non-zero.  The spatial components of the 
result are
\begin{eqnarray}
\frac{Dp^{k}}{dt}&\equiv&\frac{dp^{k}}{dt}+\Gamma^{k}_{\ ij}p^{i}\frac{dx^{j}}{dt} \nonumber \\
&=&-\left(\frac{mc^{2}}{E}\right)\frac{\partial V}{\partial x_{k}}
+q\left ({F^{k}}_{0}+\frac{v^{j}}{c}{F^{k}}_{j}\right)  ,
\label{GEOM}
\end{eqnarray}
and the time component reads
\begin{equation}\frac{DP^{0}}{dt}=\frac{d}{dt}\left(\frac{E}{c}\right)=\frac{q}{c}F^{0}_{\ j}v^{j} ,
\label{TimeComp}
\end{equation}
where $v^{j}\equiv dx^{j}/dt$, $p^{k}=mdx^{k}/d\tau$, 
$E=mc^{2}/\sqrt{1-v^{2}/c^{2}}$ (no longer a constant of the motion), 
and, as before, $m=\bar{m}\mathcal{R}$ and $q=\bar{q}$.  Note that 
the ``Newtonian gravitational three-force'' 
$F^{k}_{G}=-(mc^{2}/E)\partial V/\partial x_{k}$ vanishes as the 
particle speed approaches $c$ (as $E/mc^{2}\rightarrow \infty$), and the 
electromagnetic three-force has the Lorentz form 
[$F^{k}=q(E^{k}+e^{k}_{\ ij}v^{i}B^{j}/c)$] if and only if the electric 
and magnetic fields in G-space are defined as
\begin{eqnarray}
E^{k}&\equiv&{F^{k}}_{0}=\mathcal{R}^{3}\bar{E}^{k}  , 
\label{GE} \\
B^{k}&\equiv&\frac{1}{2}e^{k}_{\ ij}F^{ij}=\mathcal{R}^{2}\bar{B}^{k}  ,
\label{GB}
\end{eqnarray}
and so 
\begin{eqnarray}
E_{i}&=&g_{ij}E^{j}=F^{0}_{\ \ i} , \label{Elow} \\
F^{i}_{\ j}&=&e^{i\ k}_{\ j}B_{k} , \label{MagPart}
\end{eqnarray}
where $e_{kij}$ is the permutation tensor in G-space 
[$e_{kij}=\sqrt{g}\epsilon_{kij}$ and $e^{k}_{\ ij}=g^{kl}e_{lij}$, 
where $\epsilon_{ijk}$ is the permutation symbol and 
$g=det(g_{ij})$].  Thus we have

\begin{quote}

\textbf{Result X:} \textit{The equations of motion in Gaussian space 
read}
\begin{equation}
\frac{d\mathbf{p}}{dt}=-\left(\frac{mc^{2}}{E}\right)\nabla V
+q\left(\mathbf{E}+\frac{\mathbf{v}}{c}\times\mathbf{B}\right)  ,
\label{GspaceEOM}
\end{equation}
\begin{equation}
\frac{dE}{dt}=q\mathbf{E}\cdot\mathbf{v} ,
\label{EnergyEOM}
\end{equation}
\textit{where $d\mathbf{p}/dt$ is the absolute derivative in G-space of the 
three-momentum 
$\mathbf{p} =m\mathbf{v}/\sqrt{1-v^{2}/c^{2}}$  and 
the electric and 
magnetic fields in Gaussian space, $\mathbf{E}$ and $\mathbf{B}$, are related 
to the electric and magnetic fields $\bar{\mathbf{E}}$ and 
$\bar{\mathbf{B}}$ measured locally in MO-space by the conformal 
transformations}
\begin{eqnarray}
\mathbf{E}&=&\mathcal{R}^{3}\bar{\mathbf{E}} ,
\label{GspaceE} \\
\mathbf{B}&=&\mathcal{R}^{2}\bar{\mathbf{B}} .
\label{GspaceB}
\end{eqnarray}
\textit{The gravitational force $\mathbf{F}_{G}=-(mc^{2}/E)\nabla V$ 
in Gaussian space tends to zero as $v\rightarrow c$ because 
$E\rightarrow \infty$ in this limit.}

\end{quote}
Just as $\bar{\mathbf{E}}$ and $\bar{\mathbf{B}}$ are the locally 
measured fields in MO-space (measured using length and time standards 
based on a local free-running atomic clock), $\mathbf{E}$ and 
$\mathbf{B}$ are locally measured fields in the SO-picture (measured 
using time and length standards based on a local clock that is slaved 
to our single-observer's clock by time signals). Thus $\mathbf{E}$ and 
$\mathbf{B}$ are the physically correct local fields in the view of our
self-centered observer (the fields are ``corrected'' for time and space 
dilation at the location under consideration).
Using (\ref{GspaceE}) and (\ref{GspaceB})  in equations 
(\ref{MOKipMaxwells}) and expanding those equations in 
terms of G-space variables (specifically going over to G-space covariant derivatives in place of MO-space ones), we obtain the Maxwell equations in 
G-space:

\begin{quote}

\textbf{Result XI:} \textit{The Maxwell equations in Gaussian space,}
\begin{subequations}
\label{GaussMaxwells}
\begin{eqnarray}
\nabla\cdot\mathbf{E} &=& 4\pi\rho ,
  \label{SOmax1} \\*
\nabla\cdot\mathbf{B} &=& 0  ,
  \label{SOMax2} \\*
\nabla\times\mathbf{B} - 
\frac{1}{c}\frac{\partial\mathbf{E}}{\partial 
t} &=& \frac{4\pi}{c} \mathbf{j} ,
  \label{SOmax3} \\*
\nabla\times\mathbf{E} + \frac{1}{c}
\frac{\partial\mathbf{B}}{\partial 
t} &=& 0 ,
\label{SOmax4}
\end{eqnarray}
\end{subequations}
\textit{are formally identical to the Maxwell equations in the absence of a 
gravitational field; the only difference being the divergence and 
curl in these equations are the ones appropriate to the non-Euclidean G-space metric 
$g_{ij}$ instead of the flat-space metric of classical physics. The 
charge and current densities in Gaussian space are expressed in terms 
of those measured locally in the MO-picture by the conformal 
transformation laws}
\begin{eqnarray}
\rho&=&\mathcal{R}^{3}\bar{\rho} ,
\label{GspaceRoh} \\
\mathbf{j}&=&\mathcal{R}^{4}\bar{\mathbf{j}} .
\label{GspaceJ}
\end{eqnarray}

\end{quote}
The charge and current densities in G-space, equations 
(\ref{GspaceRoh}) and (\ref{GspaceJ}), are the physically correct ones 
in the view of our single observer: 
 $\rho$ is the  charge per unit 
volume as measured with the single-observer length standard and 
$\mathbf{j}$ is  $\rho$ times the velocity of this charge measured 
using single-observer time $t$.  

Equations (\ref{SOmax1})-(\ref{SOmax4}) show that, in G-space, the 
dielectric constant and magnetic permeability of the vacuum are 
unity everywhere ($\bar{\epsilon} = \bar{\mu} = 1$), and \emph{there are no 
fictitious electric or magnetic properties of empty space in this 
picture. }

\subsection{\textbf{Gauss' law, Faraday's law, and Ampere's law in 
Gaussian Space}}  

In the non-Euclidean G-space the integral theorems of Gauss and Stokes' apply in there 
usual forms:
\begin{equation}
\int_{V}\nabla\cdot\mathbf{W}dv = 
\oint_{S}\mathbf{W}\cdot d\mathbf{a},
\label{GaussLaw}
\end{equation}
\begin{equation}
\int_{S}(\nabla\times\mathbf{W})\cdot d\mathbf{a} = 
\oint_{C}\mathbf{W}\cdot d\mathbf{l},
\label{StokesLaw}
\end{equation}
where $\mathbf{W}$ is any well behaved vector field, $V$ is a volume in 
G-space (with volume element $dv = 
\sqrt{g}d^{3}x$) bounded by the closed surface $S$ in 
(\ref{GaussLaw}). In Eq.~(\ref{StokesLaw}) $S$ is an open surface with 
area element $d\mathbf{a}$ bounded by the closed contour $C$ with displacement 
element  $d\mathbf{l}$ along the contour.  

Applying these theorems to the Maxwell 
equations (\ref{GaussMaxwells}), we  
obtain the integral forms of Maxwell's equations in G-space:
\begin{subequations}
\label{Gauss3Maxwells}
\begin{eqnarray}
\oint_{S}\mathbf{E}\cdot d\mathbf{a} &=& 4\pi 
\int_{V}\rho dv,
\label{IntMax1} \\*
\oint_{S}\mathbf{B}\cdot d\mathbf{a} &=&0, 
\label{IntMax2} \\*
\oint_{C}\mathbf{E}\cdot d\mathbf{l} &=& 
-\frac{1}{c}\frac{d}{dt}\int_{S}\mathbf{B}\cdot d\mathbf{a}, 
\label{IntMax3} \\*
\oint_{C}\mathbf{B}\cdot d\mathbf{l} &=& 
\frac{4\pi}{c}\int_{S}\mathbf{j}\cdot d\mathbf{a}+                           
\frac{1}{c}\frac{d}{dt}\int_{S}\mathbf{E}\cdot d\mathbf{a}, 
\label{IntMax4}
\end{eqnarray}
\end{subequations}

Thus we have the following electromagnetic laws in Gaussian space which are 
familiar from flat-space electrodynamics:

\begin{quote}

\textbf{Result XII (Gauss' Law):} \textit{The flux of the electric field through any 
closed surface in G-space is $4\pi$ times the total charge inside the 
surface.}
\end{quote}
\begin{quote}
\textbf{Result XIII (No Magnetic Monopoles):} \textit{The flux of the magnetic field 
through any closed surface in G-space is zero.}
\end{quote}
\begin{quote}
\textbf{Result XIV (Faraday's Law):}  \textit{The electromotive force  (the line integral of 
$\mathbf{E}$ around the closed contour $C$) is $-1/c$ times 
the rate of change of the flux of $\mathbf{B}$ through any 
surface $S$ having the contour $C$ as edge.}
\end{quote}
\begin{quote}
\textbf{Result XV (Ampere's Law):} \textit{The line integral of $\mathbf{B}$ 
around a closed contour $C$ equals $4\pi/c$ times the total current 
passing through the contour, including the displacement current}
\begin{equation}
I_{D} = \frac{1}{4\pi}\frac{d}{dt}\int_{S}\mathbf{E}\cdot 
d\mathbf{a}.
\label{DispCurr}
\end{equation}

\end{quote} 

These results imply that Faraday's picture of electric and magnetic 
field lines is applicable in G-space. As in flat space, these laws are 
useful for the calculation of electric and magnetic fields when the 
symmetry of the problem is simple.  The only substantive change is that lengths of contours and areas of surfaces are calculated using the non-Euclidean G-space metric instead of the flat-space metric of classical electrodynamics.

\begin{quote}

\textbf{MO-Picture Interpretation:}  It is important to note that, generally speaking, Results XII through XV do not hold true in MO-space:
\end{quote}
\begin{itemize}
\item Result XII implies that electric field lines in G-space start on positive charges, end on negative 
charges, or go off to spatial infinity (if such exits).  That is to 
say, electric field lines do not start or end in empty space 
[this is not the case in MO-space where electric field lines can begin or end in empty space 
due to an inhomogeneous dielectric constant of the vacuum].  

\item Result XIII states that magnetic field lines neither begin nor 
end in G-space.  This means, as in Euclidean space, that either 
magnetic field lines form closed loops or else they go off to spatial 
infinity (if such exits) [this result does not apply in MO-space 
where an inhomogeneous permeability of empty space can cause 
magnetic field lines to begin or end in vacuum].  

\item Result  XIV, Faraday's Law, applies in 
MO-space as well as G-space when we employ the material Maxwell equations 
(\ref{MOMaxwells}), but does not apply in MO-space when using the 
Maxwell equations for locally measured fields 
(\ref{MOKipMaxwells}). 

\item Result XV, Ampere's Law with displacement current, does not apply in 
MO-space for either the material Maxwell equations 
(\ref{MOMaxwells}) or the Maxwell equations for locally 
measured fields (\ref{MOKipMaxwells}).
\end{itemize}

\section{\label{sec:DilationTest}A POSSIBLE DIRECT TEST OF GRAVITATIONAL SPACE DILATION}

The formalism of gravitational space dilation is based on the scaling 
laws
\begin{subequations}
\label{Scaling2}
\begin{eqnarray}
d\tau&=&\frac{d\bar{\tau}}{\mathcal{R}} ,
\label{TimeDilation} \\
d\ell&=&\frac{d\bar{\ell}}{\mathcal{R}} ,
\label{SpaceDilation}
\end{eqnarray}
\end{subequations}
relating the locally measured time $d\bar{\tau}$ and length 
$d\bar{\ell}$ to the time $d\tau$ and length $d\ell$ measured by a 
distant observer $O$.  As noted 
earlier, Eq.~(\ref{TimeDilation}) has been well tested in 
gravitational time dilation experiments.

In the present section we show that, in principle, space dilation 
[Eq.~(\ref{SpaceDilation})] can also be measured directly, and that a 
solar-system test of space dilation seems to be within the capability 
of current technology.

But hasn't space dilation already been tested in the relativistic 
radar echo delay experiment?  Placing the sun's mass near the radar 
propagation path increases the length of that path in Gaussian space 
and this nicely accounts for the additional radar delay.  Isn't this 
a direct test of gravitational length dilation?  The answer to these 
questions is in the negative.  The radar delay experiment measures the 
difference in radar path length (G-space distance) with the sun 
``near to'' and ``far from'' the radar propagation path, whereas the 
space dilation formula (\ref{SpaceDilation}) relates the proper 
lengths $d\bar{\ell}$ and $d\ell$ of a single spatial interval 
measured by local and distant observers (the proper lengths in the 
MO- and SO-pictures, respectively) in a single fixed gravitational 
field.  These are quite different concepts and should not be 
confused.  To test the space dilation formula (\ref{SpaceDilation}) 
one must measure $d\bar{\ell}$ and $d\ell$ independently and then 
determine whether or not these results are related by equation 
(\ref{SpaceDilation}).  Fortunately, the two lengths $\bar{\ell}$ 
and $\ell$ (they are, of course, finite in any experiment) can both be 
measured by means of radar.  The MO-picture proper length $\bar{\ell}$ 
is the radar distance measured locally by a radar located, say, at one 
end of the interval $\bar{\ell}$ with a transponder at the other end 
(we assume that the gravitational potential is essentially constant 
over this length).  And, as shown in Sec.~\ref{sec:RadarDelay}, the SO-proper 
length $\ell$ of the same spatial interval is measured by a radar 
located at our distant observer $O$, perhaps using transponders at 
both ends of the interval $\bar{\ell}$ (or $\ell$)

\begin{figure}
\includegraphics[width=3in]{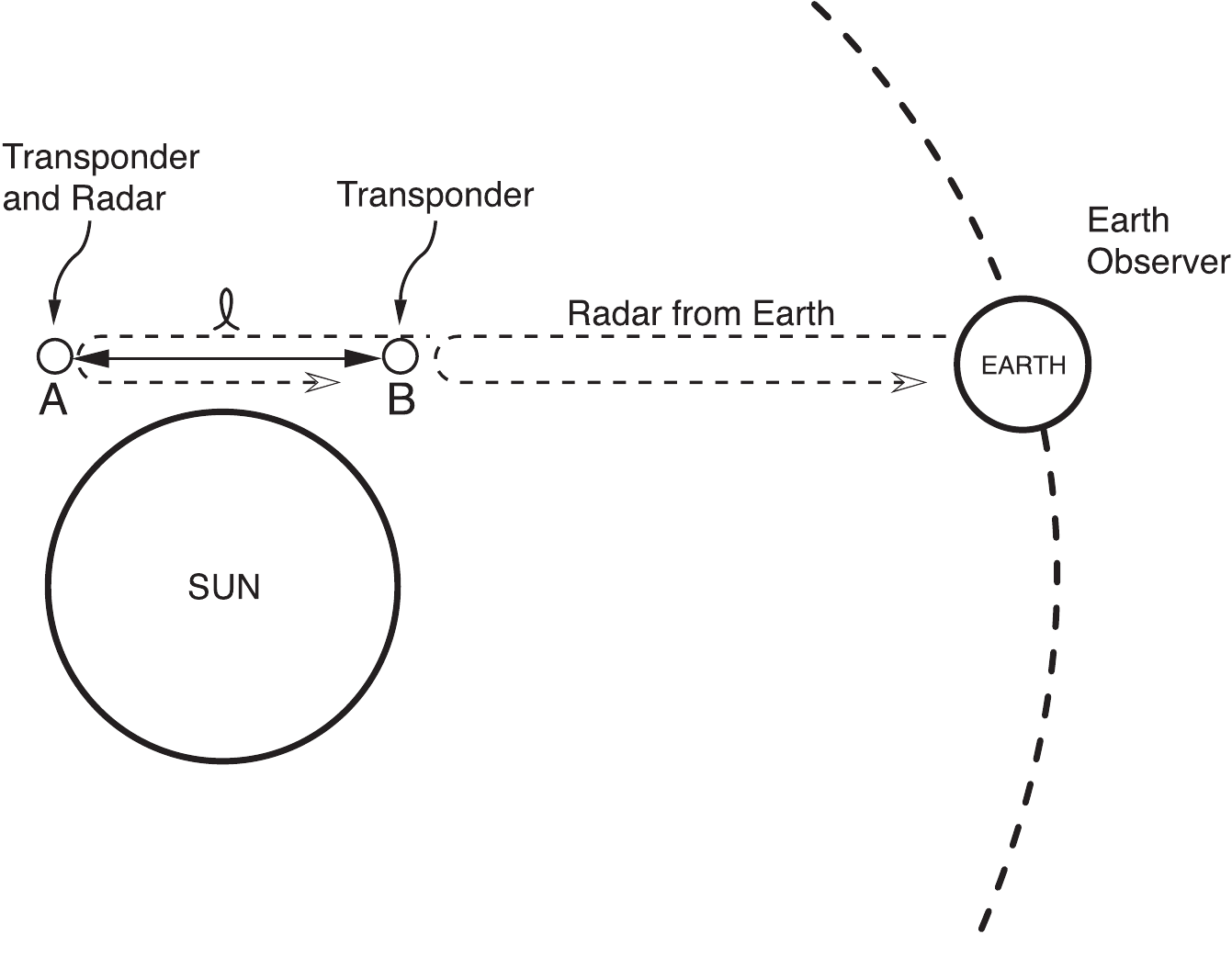}
\caption{\label{fig6}Spacecraft $A$ uses radar to locally measure the 
distance $\bar{\ell}$ to spacecraft $B$ in the sun's gravitational 
field and transmits this information to Earth.  This is the proper 
distance in the MO-picture.  Observer $O$ on Earth uses radar to 
measure the same displacement $AB$ from his distant location and 
obtains the dilated value $\ell=\bar{\ell}/\mathcal{R}$.  In this way 
the space-dilation formula (\ref{SpaceDilation}) may be tested, at 
least in principle and perhaps in practice.}
\end{figure}

In a solar-system experiment this might be accomplished as depicted 
in Fig.~\ref{fig6}.
Two spacecraft, $A$ and $B$, orbit the sun with proper distances 
$\bar{\ell}$ and $\ell$ between them in the MO- and SO-pictures, 
respectively.  $A$ and $B$ are equipped with radar transponders and $A$ 
also has an on-board radar for measuring the distance $\bar{\ell}$ to 
$B$, and a radio transmitter for sending this information to Earth.  
At the moment under consideration, $A$, $B$, and Earth lie on a 
G-space geodesic (a light ray) so that radar from earth measures the 
G-space distance $\ell$ directly.  In this way $\bar{\ell}$ and 
$\ell$ for the interval $AB$ are independently measured and one can 
check that the space dilation formula $\ell=\bar{\ell}/\mathcal{R}$ 
is satisfied.

This is, of course, an extremely idealized experiment.  In practice, 
the spacecraft $A$ and $B$ would not be lined up so neatly, and the 
Earth-based observer would measure the projection $\ell cos\theta$ of 
the length $\ell$ onto his line of sight, requiring knowledge of the 
angle $\theta$ between $AB$ and the line $OA$ extended (presumably 
this could be obtained from orbital calculations).  Then there is the 
problem of the motions of $A$, $B$, and Earth during the propagation 
of radar pulses, and so on.  Therefore, our description of the 
experiment is very crude indeed.  Nevertheless, an 
order-of-magnitude estimate based on the accuracy achieved in passed 
radar delay experiments suggests that such an experiment might be 
within the capability of existing technology.

\section{\label{sec:Applications}APPLICATIONS}

In this section we apply the single-observer formalism to a number of simple 
problems chosen to emphasize how very different physical 
interpretations can be in the MO- and SO-pictures.

\subsection{Photon Frequency}

In the MO-picture, an electromagnetic wave (or photon) propagating 
from point $A$ to point $B$ in a static gravitational field 
experiences a change in frequency from $\bar{f}_{A}$ to $\bar{f}_{B}$ 
described by the gravitational red-shift formula
\begin{equation}
\frac{\bar{f}_{A}}{\bar{f}_{B}}=
\frac{\mathcal{R}(A)}{\mathcal{R}(B)}=
\sqrt{\frac{\bar{g}_{00}(A)}{\bar{g}_{00}(B)}}  .
\label{RedShift}
\end{equation}

The fundamental scaling law $\tau=\bar{\tau}/\mathcal{R}$ for time 
intervals from the MO-picture to the SO-picture implies the scaling 
law $f=\mathcal{R}\bar{f}$ for frequencies.  Hence the gravitational 
red-shift relation (\ref{RedShift}) becomes
\begin{equation}
f_{A}=f_{B}
\label{SOredshift}
\end{equation}
in Gaussian space (these are the frequencies measured with slave-clock 
time standards at $A$ and $B$).

\begin{quote}

\textbf{Result XVI:} \textit{In the SO-picture, the frequency and 
wavelength of an electromagnetic wave (or photon) are unchanged by 
propagation in a static gravitational field.}

\end{quote}

In G-space the wavelength $\lambda=c/f$ is also unchanged because $c$ and 
$f$ are unchanged (our single observer holds the view that the 
frequency shift in the MO-picture is a spurious effect resulting from 
the use of local time standards that run at different rates at 
different points in the gravitational field, as opposed to using a 
single global time that increases at the same rate everywhere).

Because the photon frequency $f$ is unchanged by propagation in the 
SO-picture, the photon energy $E=hf$ is also unchanged.  The relevant 
conformal transformation laws are $f=\mathcal{R}\bar{f}$, $h=\bar{h}$, 
and $E=\mathcal{R}\bar{E}$, i.e., Plank's constant is the same in the 
two pictures (see Appensix A for a derivation).  The gravitational ``red shift'' in the SO-picture is 
attributed to a change in atomic rest energy with position 
[$mc^{2}=\bar{m}\mathcal{R}(\mathbf{x})c^{2}$].  Each energy level of 
the atom changes in this way with position and, therefore, an atomic 
transition that is in resonance with light of frequency $f$ at one 
point of space will not be in resonance with the same light wave when moved to a different 
location.  The change in an atomic energy level is the work done in 
lifting the atom (quasistatically) against the ``Newtonian gravitational 
force'' $\mathbf{F}=-\nabla V=-\nabla (mc^{2})$.  Hence, in the 
SO-picture, there is no ``gravitational red shift of light'' but 
instead a shift of atomic energy levels $E_{n}=m_{n}c^{2}$.  In the 
SO-picture, the position-dependent energy levels of the atom 
$E_{n}(\mathbf{x})$ show themselves in two ways: (1) the principle of 
virtual work tells us that an atom in state $n$ experiences a 
gravitational force $\mathbf{F}=-\nabla E_{n}$, and (2) the change in 
atomic transition frequencies $\omega_{nm}=(E_{n}-E_{m})/\hbar$ 
accounts for the gravitational ``red shift'' observations.

\subsection{Falling Toward a Black Hole}

Consider a particle falling radially inward from rest at $r=\infty$ 
to a spherical black hole of Schwarzschild radius $r_{s}=2m$.  At 
radial coordinate $r$ it's inward velocity on the time scale of a 
distant observer (proper distance $d\bar{\ell}$ traveled in coordinate 
time $d\bar{t}$) is
\begin{equation}
\bar{v}_{r}=-\frac{d\bar{\ell}}{d\bar{t}}=\sqrt{\frac{2mc^{2}}{r}\left(1-\frac{2m}{r}\right)} .
\label{MOradialVelocity}
\end{equation}
This well-known result in the MO-picture states that the speed of the 
falling particle first increases and then tends to zero as 
$r\rightarrow 2m$ in such a way that the particle never crosses the 
Schwarzschild sphere at $r=2m$.  It hovers just outside of this sphere 
indefinitely.  In fact, the acceleration of the particle on this time 
scale,
\begin{equation}
\bar{g}_{r}=\frac{d\bar{v}_{r}}{d\bar{t}}=
\frac{mc^{2}}{r^{2}}\left(\frac{4m}{r}-1\right)\sqrt{1-\frac{2m}{r}} ,
\label{SchwarzAccel}
\end{equation}
is radially outward ($\bar{g}_{r}>0$) for $r<4m$!  It is this 
``repulsive gravitational acceleration'' that slows the particle and 
prevents it from crossing the Schwarzschild sphere.  Such is the 
accepted description of falling-particle motion for a distant observer 
in the MO-picture.  It is a counter intuitive description if our 
intuition tells us that the gravitational acceleration ought always to 
be inward.

In the SO-picture,  the speed of the falling particle increases 
monotonically on the time scale $t=\bar{t}$ of the observer at 
infinity,
\begin{equation}
v_{r}=-\frac{d\ell}{dt}=-\sqrt{\frac{2mc^{2}}{r}}  ,
\label{SOvelocity}
\end{equation}
fortuitously having the same form as in Newtonian mechanics, with no 
outward gravitational acceleration (the velocity increases to $v_{r}=c$ 
at $r=2m$). \emph{How can this velocity of fall possibly be consistent 
with the distant observer's observation that the particle never 
reaches the Schwarzschild surface?}  The answer is that, whereas the 
proper distance from any initial radial coordinate $r=r_{0}$ to 
$r=2m$, namely
\begin{equation}
\bar{\ell}=\int_{2m}^{r_{0}}d\bar{\ell}=\int_{2m}^{r_{0}}\frac{dr}{\sqrt{1-2m/r}} ,
\label{MOdistance}
\end{equation}
is finite in the MO-picture, the same interval ($2m$, $r_{0}$) has 
infinite proper length
\begin{equation}
\ell=\int_{2m}^{r_{0}}d\ell=\int_{2m}^{r_{0}}\frac{dr}{1-2m/r}=\infty
\label{SOdistance}
\end{equation}
in the SO-picture.  So the particle never reaches the Schwarzschild 
surface in the SO-picture simply because the G-space distance to that 
surface is infinitely great.  This conclusion is consistent with  
radar measurements made from any finite radial coordinate $r=r_{0}$.  Because 
proper distance in the SO-picture is radar distance, the radar 
distance to the falling particle increases without bound as the 
particle approaches $r=2m$, and there is never a radar return from the 
Schwarzschild sphere, as one would expect for an infinitely distant 
object.

\subsection{Propagation of Photon Polarization}

In local Cartesian coordinates at an arbitrary point of G-space, the 
Maxwell equations (\ref{GaussMaxwells}) are identical in form 
to the vacuum Maxwell equations in flat space with rectangular coordinates.  In these coordinates, light travels along straight 
lines at speed $c$ (at least in geometrical-optics 
approximation), and straight-line propagation in local Cartesian 
coordinates is equivalent to geodesic propagation in G-space.  Hence, 
in the SO-picture, light is bent only by the curvature of G-space and 
not by the fictitious dielectric or magnetic properties of the 
vacuum (these contribute to the light deflection when analyzed using 
Maxwell's equations in the MO-picture).

To say that light rays follow geodesics of G-space is to say that the 
tangent unit vector to the ray $\hat{\mathbf{k}}$ is parallel propagated 
along the ray.  The unit electric polarization vector 
$\hat{\mathbf{e}}$ of a linearly polarized light ray, e.g., a laser 
beam, also undergoes parallel transport along the ray  
because it is unchanged by propagation in local Cartesian 
coordinates, and so does the magnetic polarization vector 
$\hat{\mathbf{b}}$, which is orthogonal to the other two unit 
vectors.  Therefore we have

\begin{quote}

\textbf{Result XVII:} \textit{The tangent unit vector 
$\hat{\mathbf{k}}$, the electric polarization unit vector 
$\hat{\mathbf{e}}$, and the magnetic polarization unit vector 
$\hat{\mathbf{b}}$ of a linearly polarized light ray form an 
orthonormal set of three-vectors all of which undergo parallel transport 
along the ray in Gaussian space.}

\end{quote}

In MO-space there does not appear to be any comparably simple 
three-vector picture of the propagation of polarization.

\subsection{Interferometry in a Gravitational Field}

Light beams from a coherent source $S$ travel two distinct paths (path 
$A$ and path $B$) in a static gravitational field to the point $P$ 
where they interfere.  The intensity of interference at $P$ depends 
on the phase difference of the two beams arriving there.  Because 
wavelength $\lambda$ is constant along a ray in the SO-picture, the 
phase accumulated in propagating from $S$ to $P$ is proportional 
to the path length $\ell$ in this picture  ($\phi=2\pi 
\ell/\lambda$).  Thus the 
phase difference of the interfering beams, $\Delta\phi=2\pi 
(\ell_{A} -\ell_{B})/\lambda$, is determined by the path length 
difference $\ell_{A}-\ell_{B}$ in G-space. [In MO-space the wavelength changes as the 
light propagates and the calculation of phase difference is a bit more 
complicated, and the result is \emph{not} proportional to the proper 
path length difference $\bar{\ell}_{A}-\bar{\ell}_{B}$].

\subsection{What You Calculate is What You See}

We should emphasize that the single-observer picture describes that 
which the single observers sees directly [``What You Calculate Is 
What You See'' (WYCIWYS)].  The effects of spacetime dilation are 
already included in the formalism, and it is not necessary to correct 
for these effects at the end of a calculation, as sometimes must be 
done in the MO-picture when translating locally calculated results for 
comparison with observation from a distance.  A couple of examples 
will clarify this point.

Suppose the local observer at $P$ measures a magnetic field $\bar{B}$ 
and constructs a simple clock by placing electrons in this field which, 
according to the Lorentz force law, orbit at the cyclotron frequency 
$\bar{\omega}_{c}=e\bar{B}/\bar{m}_{e}$.  For the distant observer 
at $O$, Eq.~(\ref{GspaceB}) indicates the magnetic field is 
$B=\mathcal{R}^{2}\bar{B}$, and the mass of the electron for this 
observer is 
$m_{e}=\mathcal{R}\bar{m}_{e}$ (the charge is the same in both 
pictures: $q=\bar{q}=e$).  Using the same Lorentz force law, observer 
$O$ calculates cyclotron frequency
\begin{equation}
\omega_{c}=\frac{eB}{m_{e}}=\frac{e\mathcal{R}^{2}\bar{B}}{\bar{m}_{e}\mathcal{R}}
=\mathcal{R}\bar{\omega}_{c} ,
\label{CyclotronFrequency}
\end{equation}
which is the red shifted frequency seen by this observer.  

If our single observer uses Maxwell equations in G-space to calculate 
electromagnetic phenomena, the results of the calculations are 
automatically time-scaled to what he observes (WYCIWYS). No 
additional correction for spacetime dilation is necessary (there is, 
of course, a retardation delay in observing the result due to the 
finite light propagation speed).

As a second example, consider the transition frequencies of the 
Schr\"odinger hydrogen atom, which for the local observer are
\begin{equation}
\bar{\omega}_{nm}=\frac{\bar{m}_{e}\bar{e}^{4}}{2\bar{\hbar}^{3}}\left(\frac{1}{m^{2}}-\frac{1}{n^{2}}\right)  .
\label{TransFrequency}
\end{equation}
The charge, mass, and Plank constant scale to the SO-picture as 
$e=\bar{e}$, $m_{e}=\mathcal{R}\bar{m}_{e}$, and $\hbar=\bar{\hbar}$, 
respectively.  Hence, using the same Schr\"odinger equation, observer 
$O$ calculates transition frequencies
\begin{equation}
\omega_{nm}=\frac{m_{e}e^{4}}{2\hbar^{3}}\left(\frac{1}{m^{2}}-\frac{1}{n^{2}}\right)  
=\mathcal{R}\bar{\omega}_{nm}  ,
\label{SOTransFrequency}
\end{equation}
in agreement with the gravitational red-shift formula, but only if 
the photon frequency does not change while propagating in Gaussian 
space, as deduced earlier in this section.

\subsection{Thermodynamic Equilibrium in Gaussian Space}

It is a fundamental result of classical thermodynamics that two systems in thermal contact are in thermal equilibrium when their temperatures (among other things) are equal.  For example, the atmosphere of a planet in thermal equilibrium has constant temperature throughout.

It is surprising, therefore, to learn that, in general relativity, this in \emph{not} the case.  In Tolman's classic volume \cite{Tolman}, we find that thermal equilibrium for a static fluid sphere held together by gravity is characterized by a position-dependent temperature $\bar{T}(\mathbf{x})$,
\begin{equation}
\frac{\bar{T}(\mathbf{x})}{\bar{T}(\mathbf{x}_{0})}=\sqrt{\frac{\bar{g}_{00}(\mathbf{x}_{0})}{\bar{g}_{00}(\mathbf{x})}}=\frac{1}{\mathcal{R}(\mathbf{x})} ,
\label{TempBar}
\end{equation}
where $\bar{T}(\mathbf{x}_{0})$ is the temperature at the position $\mathbf{x}_{0}$ of observer $O$. This means, for example, that the temperature of an equilibrium atmosphere in Schwarzschild geometry is larger the closer we are to the Schwarzschild surface at $r=2m$.  This is thermal equilibrium in the MO-picture.

The equilibrium temperature distribution in the SO-picture is different from that in the MO-picture.  To determine the conformal transformation law for temperature, we first note that the number of states accessible to a system $\Omega$, or the ``disorder'' of the system, as measured by the entropy
\begin{equation}
S=-k_{B}\sum_{n}P_{n}lnP_{n}=k_{B}ln\Omega
\label{Entropy}
\end{equation}
is clearly independent of the time and length standards chosen by an observer.  Therefore the Boltzmann constant must be the same in the MO and SO pictures,
\begin{equation}
k_{B}=\bar{k}_{B}.
\label{Boltzmann}
\end{equation}
Then the quantity $k_{B}T$, with dimensions of energy, necessarily transforms as
\begin{equation}
k_{B}T=\mathcal{R}(\bar{k}_{B}\bar{T}),
\label{ThermTrans}
\end{equation}
(see Appendix A for justification) 
and, in view of (\ref{Boltzmann}), the transformation law for temperature reads
\begin{equation}
T=\mathcal{R}\bar{T}.
\label{TempTrans}
\end{equation}
Thus, recalling that $\mathcal{R}(\mathbf{x}_{0})=1$, the condition for thermal equilibrium, Eq.~(\ref{TempBar}), transformed to the SO-picture becomes the following result.

\begin{quote}
\textbf{Result XVIII:}  \textit{A simple one-component fluid in hydrostatic equilibrium is in thermal equilibrium when its temperature, in the SO-picture, is constant throughout the fluid.}
\end{quote}

This result is in marked contrast to the MO-picture result (\ref{TempBar}), where thermal equilibrium is characterized by a higher temperature at lower gravitational potential.  The SO-picture returns constant temperature to its classical role as determiner of thermal equilibrium.

The equilibrium atmosphere presents a clear example of the ``What You Calculate Is What You See'' (WYCIWYS) principle of the SO-picture.  A black body at a point $P$ of low gravitational potential and in thermal equilibrium with an  atmosphere at temperature $T$, radiates as a black body of this temperature according to observer $O$ who measures the spectrum of the radiation received from the body at $P$, and concludes it is at the same temperature as the atmosphere at his location.

The conventional relativist using the MO-picture disagrees with this conclusion.  He believes that the black body at $P$ and the atmosphere at that location are hotter than the atmosphere at $O$, because $P$ is at lower gravitational potential and the system is in thermal equilibrium.  He explains observer $O$'s observations by noting that the radiation of the hot black body at $P$ is redshifted as it propagates from $P$ to $O$, and so it only appears that the black body at $P$ has the same temperature as the atmosphere at $O$.  This argument does not impress the self-centered observer at $O$ because, in his SO-picture, he knows of no such effect as the gravitational redshift of light.  He knows only  of a shift of atomic energy levels in a gravitational field.

The two pictures are physically equivalent and make the same predictions for the result of any measurement.  The pictures differ only in the position-dependent length and time standards chosen to make the measurements.

\section{\label{sec:Conclusion}SUMMARY AND CONCLUSION}

The notion of gravitational space dilation derives from a single 
observer's view that, when a distant observer's standard time interval is dilated 
by a gravitational field, his standard length must  be 
dilated as well and by the same factor. If this were not so, the distant 
observer could not understand how the local observer obtains
the invariant value $c$ for the locally measured light speed.  
The single observer, 
making ``corrections'' for time and space dilation by means of a 
conformal transformation 
[$g_{\alpha\beta}=\bar{g}_{\alpha\beta}/\mathcal{R}^{2}$ with 
$\mathcal{R}^{2}=\bar{g}_{00}(P)/\bar{g}_{00}(O)$], arrives at 
the single-observer picture explored in this paper.  

Perhaps the best way to summarize the qualitative results and 
equations of the single-observer picture is to compare  
these with the corresponding results and equations of classical (pre 
general relativity) physics.  The similarities are striking:

\begin{itemize}
\item In the single-observer picture, as in Newtonian physics, 
gravitation is represented by a force.  The gravitational 
acceleration $\mathbf{g}_{0}=-\nabla\Phi$ of non-relativistic 
particles ($v<<c$ or $E\approx \bar{m}c^{2}$) is derivable from a 
potential $\Phi$ that obeys the same \emph{linear} Poisson equation,
\begin{equation}
\nabla^{2}\Phi=4\pi G\rho_{g}  ,
\label{ClassicalPoisson}
\end{equation}
as in Newtonian theory (an \emph{exact} fully relativistic result).

\item In Gaussian space, as in the classical Euclidean space, the 
electromagnetic three-force takes the Lorentz form,
\begin{equation}
\mathbf{F}=q\left(\mathbf{E}+\frac{\mathbf{v}}{c}\times\mathbf{B}\right)  ,
\label{Lorentz}
\end{equation}
with electric and magnetic fields that obey the vacuum Maxwell 
equations 
\begin{subequations}
\label{FinalMaxwells}
\begin{equation}
\nabla\cdot\mathbf{E} = 4\pi\rho ,
  \label{SOmax1Sum}
\end{equation}
\begin{equation}
\nabla\cdot\mathbf{B} = 0  ,
  \label{SOmax2Sum}
\end{equation}
\begin{equation}
\nabla\times\mathbf{B} - 
\frac{1}{c}\frac{\partial\mathbf{E}}{\partial 
t} = \frac{4\pi}{c} \mathbf{j} ,
  \label{SOmax3Sum}
\end{equation}
\begin{equation}
\nabla\times\mathbf{E} + \frac{1}{c}
\frac{\partial\mathbf{B}}{\partial 
t} = 0 ,
\label{SOmax4Sum}
\end{equation}
\end{subequations}
of the same form as in Maxwell's original flat-space theory 
(an \emph{exact} relativistic result).

\item  In Gaussian space, as in the classical three space, light 
rays propagate along geodesics of the three-space geometry and the 
frequency of light is unaffected by propagation in a static 
gravitational field.

\item In the single-observer picture, the energy, momentum, and 
Lagrangian for a particle in a static gravitational field,
\begin{eqnarray}
E&=&\frac{mc^{2}}{\sqrt{1-v^{2}/c^{2}}} , \label{Engy} \\
\mathbf{p}&=&\frac{m\mathbf{v}}{\sqrt{1-v^{2}/c^{2}}} , \label{Mom} \\
L&=& -mc^{2}\sqrt{1-v^{2}/c^{2}} , \label{Lag}
\end{eqnarray}
have the same forms as for a free particle in special relativity.

\item As in classical physics, thermal equilibrium in the single-observer picture is characterized by uniform temperature.
\end{itemize}

\

These results justify our referring to the single-observer picture
 as the \emph{``classical picture''} in general relativity.  It is far 
 closer in spirit to classical physics than the usual many-observer 
 picture that is traditionally used in general relativity.  Of course, the 
 latter picture is not at all incorrect in its predictions (the two 
 pictures are physically equivalent), but the classical picture is 
 likely to be of interest to those who prefer to lean on their 
 classical intuition when interpreting the results of general 
 relativity.
 
 There are, of course, differences between the ``classical picture'' in 
 general relativity and classical physics prior to general 
 relativity:

 \begin{itemize}
 \item  In the \emph{classical picture}, the rest mass 
 $m=\bar{m}\mathcal{R}(\mathbf{x})$ and the rest energy $E=mc^{2}$ of 
 a particle are position dependent in a gravitational 
 field.  The rest energy is the gravitational potential energy in 
 this picture, and it's gradient is the gravitational force.
 
 The position dependence of the rest energies 
 $E_{n}=m_{n}(\mathbf{x})c^{2}$ of an atom (the atomic energy levels) 
 shows itself through the position-dependent transition frequencies 
 $\omega_{nm}=(E_{n}-E_{m})/\hbar$ which account for the 
 ``gravitational redshift'' of spectral lines (if the rest masses of 
 particles were not position-dependent, cyclotron clock frequencies 
 and atomic transition frequencies would not correctly display the 
 gravitational time dilation effect in this picture).
 
 \item  When a particle moves fast ($v\sim c$ or $E>\bar{m}c^{2}$) 
 it's acceleration in Gaussian space,
 \begin{equation}
 \mathbf{g}=\left(\frac{\bar{m}c^{2}}{E}\right)^{2}\mathbf{g}_{0} ,
 \label{GaussAccel}
 \end{equation}
 decreases from the Newtonian value $\mathbf{g}_{0}=-\nabla\Phi$ by 
 the factor $(\bar{m}c^{2}/E)^{2}$ and approaches zero (geodesic 
 motion in Gaussian space) as $E\rightarrow \infty$.  Thus slow 
 particles experience the Newtonian gravitational acceleration 
 $\mathbf{g}_{0}$, but ultra-relativistic particles (and the photon) 
 experience no acceleration and travel on geodesics of Gaussian space.
 
 \item  Perhaps most importantly, Gaussian space is curved 
 (non-Euclidean), whereas the three-space of classical physics is 
 flat.  In the single-observer picture, three of the four classic 
 tests of general relativity are attributed to the curvature of 
 Gaussian space.  In this picture, the precession of perihelia, the 
 bending of light by the sun's gravitational field, and the 
 relativistic radar echo delay are all measures of the curvature of 
 the the solar G-space, and these interpretations are exact and 
 complete (as opposed to an interpretation based on a limited number 
 of terms in a perturbation expansion).  This economy of 
 interpretation seems desirable.
\end{itemize}

Another striking feature of the single-observer picture is the 
independence of all electromagnetic 
phenomena, from the Newtonian gravitational field 
$\mathbf{g}_{0}=-\nabla\Phi$!  Electric and magnetic fields are 
``distorted'' from their classical values by the non-Euclidean G-space 
geometry (and possibly by a non-classical topology of this space), 
but neither the potential $\Phi$ nor it's gradient $\mathbf{g}_{0}=-\nabla\Phi$
appear in the Maxwell equations (\ref{FinalMaxwells})
in G-space, and consequently these have no affect on the fields 
$\mathbf{E}$ and $\mathbf{B}$ for any prescribed G-space metric 
$g_{ij}$.  This is perhaps most clearly evident in local Cartesian 
coordinates where slow material particles fall with acceleration due 
to gravity $\mathbf{g}_{0}=-\nabla\Phi$, but the Maxwell equations 
(\ref{FinalMaxwells}) generate and propagate electric 
and magnetic fields in exactly the same manner as when $\mathbf{g}_{0}=0$.
This result seems to be at variance with Einstein's original 
principle-of-equivalence argument for the bending of light by the 
``gravitational field'' in an upward accelerating elevator.  But we 
must remember that Einstein's argument gives only half the correct light 
deflection in the sun's gravitational field, as in Einstein's 
 calculation \cite{Einstein1} prior to general relativity.  In the 
SO-picture, no part of the light deflection is attributed to the 
principle of equivalence in this way.  The deflection is entirely due 
to the curvature of the Gaussian three-space.

Finally, it is noteworthy that, in Gaussian space, the results of 
calculations made with the Maxwell equations or particle equations of 
motion for phenomena at some distance from the observer $O$ require 
no correction for gravitational time dilation at the location of the 
phenomena.  The temporal scale of happenings anywhere in Gaussian space is as 
observed from $O$ (with, of course, a retardation delay due to the 
finite propagation speed $c$ of the optical image from phenomenon to 
observer); the correction for gravitational time dilation being 
already included in the formalism by means of the conformal 
transformation.

The results of the present paper (Part I) are limited to static gravitational 
fields.  In a second paper (Part II) we shall extend the 
single-observer picture (or ``space dilation'' picture) to the time-dependent cosmological metric.

\appendix
\section{Conformal Scaling Rules}

Many physical quantities may be thought of as products of mass ($M$), 
length ($L$), time ($T$),  charge ($Q$), and absolute temperature ($K$).  Under the conformal 
transformation $ds^{2}=\mathcal{R}^{2}d\bar{s}^{2}$ from the MO-picture to the SO-picture, these quantities transform as
\begin{eqnarray}
M&=&\mathcal{R}\bar{M} , \label{MScale} \\
L&=&\bar{L}/\mathcal{R} , \label{LScale}  \\
T&=&\bar{T}/\mathcal{R} , \label{TScale}  \\
Q&=&\bar{Q} .\label{QScale} \\
K&=&\mathcal{R}\bar{K}
\label{KScale}
\end{eqnarray}
Therefore, if an observable $X$ has dimensions 
$[X]=M^{n_{M}}L^{n_{L}}T^{n_{T}}Q^{n_{Q}}K^{n_{K}}$ we may think of it as 
composed of $n_{M}$ factors of mass, $n_{L}$ factors of length, 
$n_{T}$ factors of time,  $n_{Q}$ factors of charge, and $n_{K}$ factors of temperature, in which case 
we would expect the quantity to transform under the conformal 
transformation in the same way as the quantity
$M^{n_{M}}L^{n_{L}}T^{n_{T}}Q^{n_{Q}}K^{n_{K}}$, namely
\begin{equation}
X=\mathcal{R}^{n_{M}+n_{K}-n_{L}-n_{T}}\bar{X} 
\label{ScalingLaw}
\end{equation}
(here we are using $M$, $L$, $T$, $Q$, and $K$ both as symbols for a particular mass, 
length, time , charge, and temperature and as indicators of the dimensions of these 
quantities).  We will refer to quantities that transform in this way 
as \emph{fundamental} observables.

The metric $g_{\alpha\beta}$ is a fundamental observable if we 
follow the convention that all coordinates $x^{\mu}$ are dimensionless.
The metric tensor then has dimensions of length squared, so 
that $ds$ ($ds^{2}=g_{\alpha\beta}dx^{\alpha}dx^{\beta}$) has 
dimensions of length for any coordinate displacement $dx^{\alpha}$, 
and $g_{\alpha\beta}=\bar{g}_{\alpha\beta}/\mathcal{R}^{2}$ follows the 
pattern (\ref{ScalingLaw}).  The inverse metric $g^{\alpha\beta}$ 
then has dimensions of inverse length squared and 
$g^{\alpha\beta}=\mathcal{R}^{2}\bar{g}^{\alpha\beta}$ also follows the 
rule (\ref{ScalingLaw}).  Such a convention is in keeping with the 
view that a coordinate displacement $dx^{\alpha}$ has no length until 
a metric is specified. This convention, together with rule 
(\ref{ScalingLaw}), also implies that coordinates are unaffected by a 
change of picture ($x^{\mu}=\bar{x}^{\mu}$), and so we use the same 
coordinate symbol $x^{\mu}$ in both pictures because there is no 
need to make a distinction. 

The only caution in applying the scaling rule (\ref{ScalingLaw}) is 
that tensor components such as $P^{\mu}$ or $P_{\mu}$ often do not have the 
dimensions of the physical observable they represent.  For example, 
$P^{\mu}=mdx^{\mu}/d\tau$ has dimensions $[P^{\mu}]=M/T$ and $P_{\mu}$ 
dimensions $[P_{\mu}]=[g_{\mu\nu}P^{\nu}]=ML^{2}/T$, neither of which 
are the dimensions of mass times velocity.  Only the physical 
magnitude $P=(P_{\mu}P^{\mu})^{1/2}$ has the dimensions 
$ML/T$ of momentum.  We are keeping factors of $c$ and $G$ in all equations, rather than setting these to unity, in order to make application of the scaling rule (\ref{ScalingLaw}) more convenient.

The convention that coordinates $x^{\mu}$ are dimensionless is 
convenient also because it allows all components of the electromagnetic 
field tensor $F^{\mu}_{\;\;\nu}$ to have the same dimensions, as can 
be seen from the Lorentz formula
\begin{equation}
\frac{DP^{\mu}}{d\tau}=\frac{q}{c}F^{\mu}_{\;\;\nu}\frac{dx^{\nu}}{d\tau} .
\label{LorentzDP}
\end{equation}
The dimensions of $F^{\mu}_{\;\;\nu}$ are 
$[F^{\mu}_{\;\;\nu}]=ML/T^{2}Q$, and so the transformation rule is 
$F^{\mu}_{\;\;\nu}=\mathcal{R}^{2}\bar{F}^{\mu}_{\;\;\nu}$.  This, 
together with the transformation rule 
$J^{\mu}=\mathcal{R}^{4}\bar{J}^{\mu}$ for $J^{\mu}$, which follows from the 
dimensions $[J^{\mu}]=Q/L^{3}T$ of $J^{\mu}$, are the basis of the 
proof that Maxwell's equations are conformally invariant.

\begin{table}
\begin{center}
\begin{tabular}{|c|c|} \hline
PHYSICAL & SCALING \\
CONSTANT & LAW \\  \hline\hline
Elementary Charge  $e$& $e=\bar{e}$ \\ \hline
Speed of Light  $c$& $c=\bar{c}$ \\ \hline 
Plank's Constant $\hbar$& $\hbar=\bar{\hbar}$ \\ \hline
Gravitational Constant $G$& $G=\bar{G}/\mathcal{R}^{2}$ \\ \hline
Boltzmann Constant $k_{B}$ & $k_{B}=\bar{k}_{B}$ \\ \hline
\end{tabular}
\end{center}
\begin{quote}
\textbf{Table 1:} Conformal scaling laws for fundamental constants.
\end{quote}
\end{table}

The scaling law (\ref{ScalingLaw}) has a number of immediate 
consequences for the scaling of fundamental ``constants'' under the transformation from the MO- to the SO-picture..  Table 1 contains the scaling laws for some of 
these constants.  We see that the elementary charge, the speed of light, Plank's constant, and Boltzmann's constant are invariant under this transformation, but the gravitational "constant" changes.
The scaling rule (\ref{ScalingLaw}) does not say that all 
observables scale in this way.  Only the so-called ``fundamental'' 
quantities obey this rule.  If the definition of a quantity involves derivatives of fundamental quantities, then that quantity does not transform according to the scaling law (\ref{ScalingLaw}), although its transformation law is still easily derived.  The connection coefficients $\Gamma^{\mu}_{\;\;\alpha\beta}$ and the curvature tensor, for example, 
transform differently from (\ref{ScalingLaw}) under a conformal transformation .


\begin{thebibliography}{99}
\bibitem{Remark}  Any conformal transformation 
$ds^{2}=d\bar{s}^{2}/\mathcal{R}^{2}$ from the MO metric $d\bar{s}^{2}$ to 
a new metric $ds^{2}$ may be interpreted as a change in the rates of 
the clocks used to define the standards of time and length, where 
$\mathcal{R}=d\bar{\tau}/d\tau$ is the rate of the old standard 
clocks as measured with the new ones.
\end{thebibliography}
\end{document}